\documentclass[aps,prb,twocolumn,superscriptaddress,showpacs,showkeys]{revtex4-1}
\usepackage{amsmath,amssymb}
\usepackage{graphicx}
\usepackage{txfonts}
\usepackage{color}
\begin{document}

\title{Rotation of a single acetylene molecule on Cu(001) by tunneling electrons in STM}
\author{Yulia E. Shchadilova}
\affiliation{A. M. Prokhorov General Physical Institute, Russian Academy of Science, Moscow, Russia}
\author{Sergei G. Tikhodeev}
\email[]{tikh@gpi.ru}
\affiliation{A. M. Prokhorov General Physical Institute, Russian Academy of Science, Moscow, Russia}
\affiliation{Division of Nanotechnology and New Functional Material Science, Graduate School of Science and Engineering,
University of Toyama, Toyama, 930-8555 Japan}
\author{Magnus Paulsson}
\affiliation{Department of Physics and Electrical Engineering,
Linnaeus University, 391 82 Kalmar, Sweden}
\affiliation{Division of Nanotechnology and New Functional Material Science, Graduate School of Science and Engineering,
University of Toyama, Toyama, 930-8555 Japan}
\author{Hiromu Ueba}
\affiliation{Division of Nanotechnology and New Functional Material Science, Graduate School of Science and Engineering,
University of Toyama, Toyama, 930-8555 Japan}
\date{June 25, 2013}

\begin{abstract}
We study the elementary processes behind one of the pioneering works on STM controlled reactions of
single molecules  [Stipe et al., Phys. Rev. Lett. {\bf 81}, 1263 (1998)]. Using the Keldysh-Green function
approach for the vibrational generation rate in combination with DFT calculations to obtain realistic parameters
we  reproduce the experimental rotation rate of an acetylene molecule on a Cu(100) surface
as a function of bias voltage and tunneling current. This combined approach allows us to identify the reaction
coordinate mode of the acetylene rotation and its anharmonic coupling with the C-H stretch mode. We show that
three different elementary processes, the excitation of C-H stretch, the overtone ladder climbing  of the hindered rotational mode,
and  the combination band excitation together explain the rotation of the acetylene molecule on Cu(100).
\end{abstract}

\maketitle

Tunneling electrons from a scanning tunneling microscope (STM) form an atomic source of
electrons for electronic and vibrational excitations which can be used to manipulate individual atoms and molecules in a
controlled manner~\cite{Eigler1991,Mo1993,Stipe1997,Stipe1998b}.
The study of acetylene C$_2$H(D)$_2$ rotation on the Cu(001) surface~\cite{Stipe1998a}
was the first comprehensive and systematic experiment on a single adsorbate manipulation
made in combination with STM inelastic electron tunneling spectroscopy (STM-IETS).
This method has been established as an indispensable experimental method to gain
insight into the vibrationally mediated motions and reactions of single molecules with
STM (see, e.g., Ref.~\onlinecite{Ueba2011} and references therein).

The observed in Ref.~\onlinecite{Stipe1998a} rotation yield per
electron as a function of bias voltage for C$_2$H$_2$(D$_2$) exhibits a threshold
at 358 (266) mV. This corresponds to the excitation of the CH(D) stretch
mode which is not a rotational mode.
In this respect it is very different from, e.g.,  the rotation of a single oxygen molecule on  Pt(111)
surface~\cite{Stipe1998b}, where the hindered rotational mode can be directly excited
by tunneling electrons~\cite{Teillet-Billy2000}.
That the onset of a molecular motion/modification is caused by a vibrational mode that is
not directly responsible for the observed motion is not unique. Many other examples
have been established, e.g., the migration of CO on Pd(110)~\cite{Komeda2002}.
The rotation of C$_2$H(D)$_2$/Cu(001)~\cite{Stipe1998a} also demonstrates several
peculiar features which have not been previously understood. In this paper
we revisit the experimental findings of Ref.~\onlinecite{Stipe1998a}. Based on DFT calculations
and the Keldysh diagram technique we clarify the vibrational modes
and elementary physical processes behind the characteristic features of the
acetylene~\footnote{Rotation of C$_2$D$_2$ on Cu(001) will be
analyzed elsewhere, it can be done analogously to the current approach.} rotation on Cu(001).

\begin{figure}[b]
\begin{centering}
\includegraphics[width=0.98\columnwidth]{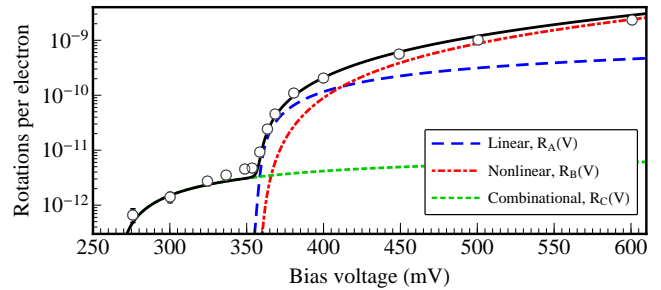}
\par\end{centering}
\caption{\label{fig:YvsV}
Rotation yield per electron as a function of bias voltage at fixed tunneling current $I=40$~nA.
Circles are experimental data from Ref.~\onlinecite{Stipe1998a}. Blue dashed line is the
linear process A, Eq.(\ref{Eq:Ra}), red dash-dotted line is the nonlinear process B, Eq.(\ref{Eq:Rb}),
and green dotted line is the combination band process C, Eq.(\ref{Eq:Rc}).
Black solid line corresponds to a sum of all processes.
Lines are calculated via Eqs.~(\ref{RABC},\ref{Eq:Ra},\ref{Eq:Rb}) with $A=2.5\times10^{-6},\:B=6\times 10^{-16}\:\textrm{s}$,
$C=1.2\times10^{-8}$,
and $\Gamma_\mathrm{iet}(\Omega_1)$,  the CH mode generation rate Eq.(\ref{Eq:Giet}), is calculated via the
Keldysh technique~\cite{Tikhodeev2004}, the  parameters are specified in the text.
}
\end{figure}

\begin{figure}[t]
\begin{centering}
\includegraphics[width=0.95\columnwidth]{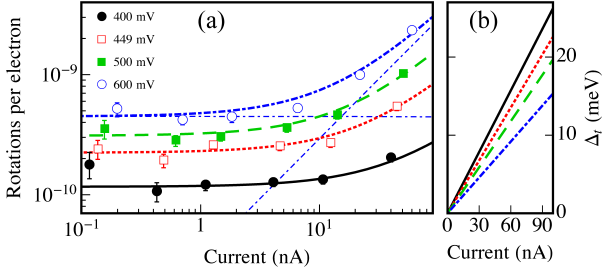}
\end{centering}
\caption{
\label{fig:YvsI}
Rotation yield per electron $Y$ as a function of tunneling current (panel a).
Symbols show the experimental data from Ref.~\onlinecite{Stipe1998a}.
Lines are the calculated results for different bias voltages: $V=400$ (black solid line), 449 (red dashed), 500
(green dash-dotted), and 600~mV (blue dotted).
Thin dash-dotted lines are the linear and  nonlinear contributions to the rotation yield  for $V=600$~mV.
The calculations are done with the same parameters as
in Fig.\ref{fig:YvsV}.  The dependencies of the
hybridization parameter $\Delta_t$ on current for different bias voltages are shown in panel b.
}
\end{figure}

The experimental data~\cite{Stipe1998a} on the C$_2$H$_2$/Cu(001) rotation  is summarized
as symbols in Figs.~\ref{fig:YvsV} and \ref{fig:YvsI}(a).
The higher threshold at 358~meV, see Fig.~\ref{fig:YvsV}, corresponds to
excitation of the CH stretch mode. In addition, a lower threshold with a much smaller
rotation yield is clearly seen at $\sim 240$~meV, we note that this does not correspond to
any vibrational energy of the C$_2$H$_2$/Cu(001), see below.
The third feature is the crossover from a single to two-electron
process with increasing tunneling current above 10~nA, see Fig.~\ref{fig:YvsI}(a).  This crossover
cannot be attributed to a resonant inelastic electron-molecule scattering with rotational excitation
as in the case of O$_2$/Pt(111)~\cite{Teillet-Billy2000}, because of the evident absence of direct excitation
of the hindered rotational modes. Moreover, the crossover cannot be attributed to
coherent ladder climbing since the reaction order (number of
electrons needed) does not approache one at high bias~\cite{Salam1994,Stipe1998b}.

The identification of the precursor to rotation is not straightforward since the molecule has a number
of low frequency vibrational modes. The hindered rotational mode of C$_2$H$_2$ on Cu(001) $\sim\ 28$~meV,
whereas the thermally measured barrier for rotation is relatively high $\epsilon_B = 169\pm3$~meV~\cite{Lauhon1999a}.
For this low-energy rotational mode to directly induce rotations,
a  multi (5-6)-electron processes is expected, whereas in the experiment, as explained above,
only a two-electron process is observed at larger tunneling currents.

\begingroup
\squeezetable
\begin{table}[h]
\caption{\label{tab:VibModes}
Calculated/experimental vibrational
energies, damping rates, and angular momentum of C$_2$H$_2$ on Cu(001).
 The frustrated rotational modes along (001) are emphasized in bold.}
\begin{ruledtabular}
\begin{tabular}{c c c c c c c c c c}
$\nu$ & Mode
    & \multicolumn{4}{c}{$\hbar\Omega_{\nu}$}
                                                            & $\gamma_\textrm{eh}^{(\nu)}$
                                                                     & $L_x^{(\nu)}$
                                                                                        & $L_y^{(\nu)}$
                                                                                                    & $L_z^{(\nu)}$ \\
&  &  \multicolumn{4}{c}{meV}     &      $10^{12}$s$^{-1}$ &       \multicolumn{3}{c}{(rel.u.) }    \\
  \hline
&     &  \multicolumn{2}{c}{theory}     &     \multicolumn{2}{c}{expt}      &  \multicolumn{4}{c}{theory}  \\
    &          &  this
                         &
                                      &EELS
                                                &  IETS \\
    &
             &work
                         &  Ref.~\onlinecite{Olsson2002}
                                      & Ref.~\onlinecite{Marinova1987}
                                                &  Ref.~\onlinecite{Stipe1998a}   &  \multicolumn{4}{c}{this work}    \\
\hline
1   & CH stretch sym            & 371               & 379           & 364                   &  358              & 1.0               &      -0.01      &    0.01          &     0      \\
2   & CH stretch asym             & 368              & 375           & 357                   & 358               & 0.7               &     1.66        &   -1.66          &     0      \\
3    &  CC stretch          & 167              & 171           & 164                   &  N.O.$^{**}$              & 2.2              &     0              &     0             &     0      \\
4    &    CH in-plane bend      \\
      & or wag, asym      & 131               & 132           & 141                   &  N.O.              & 0.2               &      1.16       &   -1.16          &     0      \\
5     &   CH in-plane bend  \\
      & or scissor, sym      & 111               & 117           & 118                  &  N.O.             & 1.5            &       0            &     0               &     0      \\
\textbf{6}  &  \textbf{CH out-of-plane bend} \\
                &  \textbf{or asym rotation} & \textbf{100} & \textbf{101}  &  \textbf{78}     &  N.O.             &\textbf{ 0.7}  & \textbf{0}      &   \textbf{0}     &  \textbf{-1.06}           \\
7   &    out-of-plane bend \\
    & or cartwheel        & 71                & 75                & N.O.                & N.O.               & 0.2                &   -0.24        &  -0.24            &  0         \\
8    &   in-plane bend or wag        & 58                & N.P.$^*$             & N.O.                & N.O.               & 2.0                &    1.26        &   -1.26            &     0      \\
9    &    molecule-Cu stretch       & 50                & N.P.            & 52                    & N.O.              &0.05              &    0              &     0             &     0      \\
10    &    in-plane rotation      & 29                & N.P.             &  N.O.               &  N.O.             &1.8             &      -0.35      &    0.35         &      0     \\
\textbf{11} & \textbf{out-of-plane rotation} & \textbf{28}   &  N.P.            &  N.O.                & N.O.             &\textbf{0.2}    &\textbf{0}       &    \textbf{0}  &   \textbf{1}       \\
12    &     out-of-plane bend     & 23                &  N.P.          & N.O.                &  N.O.             &0.04               &   -0.04         &     -0.05       &     0.03    \\
\hline
 \multicolumn{10}{l}{$^*$   N.P.: Not Published}\\
  \multicolumn{10}{l}{$^{**}$  N.O.: Not Observed}\\
\end{tabular}
\end{ruledtabular}
\end{table}
\endgroup

To analyse the mechanisms of rotation of a single C$_2$H$_2$/Cu(001) by
tunneling electrons we performed DFT calculations on a 4$\times$4 Cu(001) surface with one
adsorbed acetylene molecule. The calculations of relaxed geometries, vibrational energies, and
electron-hole pair damping rates were carried out with SIESTA~\cite{Soler2002,Frederiksen2007}.
Other details of the DFT calculations are presented in the Supplementary material, Sec.~I (\textit{SM-I} in what follows).
We also estimated the energy barrier $\epsilon_B$  for C$_2$H$_2$  rotation on Cu(001) to 100 meV using the
nudged elastic band method.

The results  for  the vibrational
energies $\hbar \Omega_\nu$ and  electron-hole damping rates $\gamma_{\mathrm{eh}}^{(\nu)}$, $\nu = 1, \ldots 12$
are given in Tab.~\ref{tab:VibModes},
in comparison with the previous theoretical~\cite{Olsson2002} and experimental~\cite{Marinova1987,Stipe1998a} results.
The last three columns of Tab.~\ref{tab:VibModes} show (in relative units) the components
of each modes angular momentum
$\vec{L}^{(\nu)} = \sum_i \Omega_\nu (\vec{r}_i - \vec{r}_\mathrm{C.M.}) \times m_i \delta\vec{r}_i^{(\nu)}$,
where $\vec{r}_i$, $m_i$ are the  atomic positions (in the acetylene molecule) and masses, $\vec{r}_\mathrm{C.M.}$ the center-of-mass,
and $\delta\vec{r}_i^{(\nu)}$ the atomic displacements in the vibrational mode $\nu$.
The high frequency modes (\# 1 \& 2 in Tab.~\ref{tab:VibModes})  are the symmetric and antisymmetric CH stretch modes.
The higher threshold energy of the rotation yield $Y(V) = R(V)/I(V)$ at $358$~mV and the corresponding peak in the
$\Delta{\rm log}(Y)/\Delta V$  plot, observed in Ref.~\onlinecite{Stipe1998a}, indicate that inelastic excitation by tunneling electrons
of these modes is a trigger for rotation. Along the reaction pathway there are only two frustrated rotation modes which has angular
momentum in the (001) direction, \# 6 and \# 11.

The full Hamiltonian of the system can be written as a sum of the electronic and vibrational (phonon) parts,
\begin{equation}\label{Eq:Hfull}
H = H_{e}(\lbrace \delta q_{\nu}\rbrace) + H_{ph},
\end{equation}
where the electronic part $H_e$ depends on the normal coordinates of the molecule $\lbrace \delta q_{\nu}\rbrace$.
Using the Newns-Anderson type Hamiltonian~\cite{Newns1969}, we write
\begin{equation}
\label{eq:He}
H_e=\varepsilon_a (\lbrace \delta q_{\nu}\rbrace)c_a^{\dagger}c_a +
 \sum_{ j= t,s} \varepsilon_{j} c_{j}^{\dagger} c_{j}+
 \sum_{j= t,s} V_{j} \left( c_j^{\dagger} c_a+h.c.\right),
\end{equation}
where the indices $s$($t$) and $a$ denote a substrate (tip) and the adsorbate, respectively; the corresponding
energy levels are $\varepsilon_{s(t)}$ and $\varepsilon_{a}(\lbrace \delta q_{\nu}\rbrace)$.
Electronic tunneling matrix elements $V_{t}$
(tip-adsorbate) and $V_{s}$ (substrate-adsorbate) give rise to a stationary tunneling current between the tip
and the substrate through the adsorbate orbital at applied bias voltage $V$.
The electron occupation functions in the substrate and tip are assumed to be
Fermi  distributions with the same temperature $T$ but
different chemical potentials $\mu_t$ and $\mu_s$,
 $\mu_s-\mu_t = eV$.

In order  to clarify the interaction parts of the full Hamiltonian~(\ref{Eq:Hfull})  behind the
experimental results on the acetylene rotation, it is convenient to
split the rotation rate into three partial processes,
\begin{equation}\label{RABC}
R(V) = R_A(V)+R_B(V)+R_C(V),
\end{equation}
where the rates $R_A(V)$ and $R_B(V)$ are, respectively, the one- and two-electron processes with a higher threshold $V\sim358$~mV,
and $R_C(V)$ is the one-electron process with a lower threshold $V\sim240$~mV.

To describe the generation of high frequency CH stretch modes, the adsorbate orbital energy in the 1st term of Eq.~(\ref{eq:He})
can be expanded~\cite{Persson1980} to the first order as
\begin{equation}
\label{eq:Ea}
\varepsilon_a(\lbrace \delta q_{h}\rbrace) \approx \varepsilon_a(0) + \chi (b_{h}^{\dagger} +b_{h}),
\end{equation}
where $\chi$ is an electron-phonon constant and $ \varepsilon_a(0)$ the unperturbed
adsorbate energy,
$b_{h}$ is the annihilation operator of the vibrational mode
$\left( \delta q_h=2^{-1/2}(b^\dagger_h+b_h) \right)$ with the frequency $\Omega_{h}=358$~meV,
which is directly excited by the inelastic tunneling current~\footnote{The energies of the high frequency CH stretch
modes are slightly different, see in Table~\ref{tab:VibModes}; we use the experimental value
from Ref.\onlinecite{Stipe1998a} in the estimates below.}.

The high frequency vibration generation rate then reads~\cite{Tikhodeev2004}
\begin{equation}\label{Eq:Giet}
\Gamma_{\mathrm{iet}}(\Omega_{h},V)=\int d\omega \rho^{(h)}_\mathrm{ph}(\omega)\Gamma_{\mathrm{in}}(\omega,\Omega_{h},V),
\end{equation}
where $\rho^{(h)}_\mathrm{ph}(\omega)=\pi^{-1} \gamma_\mathrm{eh}^{(h)}
\left[\left(\omega-\Omega_h\right)^2+ \left(\gamma_\mathrm{eh}^{(h)}\right)^2\right]^{-1}$,
$ \gamma_\mathrm{eh}^{(h)}$ is the inverse lifetime of the phonon mode $h$ due to electron-hole pair excitation, given in Tab.~\ref{tab:VibModes}.
At $T=0$~\cite{Gao1997}
\begin{equation}
\Gamma_{\mathrm{in}}(\omega, \Omega_{h}, V)\simeq \frac{\gamma_\mathrm{eh}^{(h)}}
{\hbar \Omega_{h}} \frac{\Delta_t}{\Delta_s}\left(\left|eV\right|-\hbar\omega\right)\Theta\left(\frac{\left|eV\right|}{\hbar\omega}-1\right).
\end{equation}

Our next task is to find the mechanisms of the energy transfer between the high frequency
CH stretch  mode and the reaction coordinate (RC) hindered rotational mode.
The first possibility is the direct over-barrier rotation excitation due to an inelastic
tunneling generation of the stretch mode $\Omega_{h}$~\cite{Komeda2002,Persson2002}.
The rotation rate corresponding to such single-electron process can be described
~\cite{Kumagai2012} as a linear function of the CH stretch mode generation rate,
\begin{equation}
\label{Eq:Ra}
R_A(V)=A\Gamma_\mathrm{iet}(\Omega_h,V).
\end{equation}
A good fit to the experimental data is given by $A=2.5\times10^{-6}$, see Fig.~\ref{fig:YvsV}..
The physical meaning of $A$ is the probability of a CH stretch vibration to excite the
C$_2$H$_2$ rotation over the barrier.
This value is in reasonable agreement with the well established case of  the migration of CO on Pd(110)~\cite{Komeda2002,Ueba2012}.
However, this mechanism cannot describe the two-electron process. In principle, the direct
over-barrier rotation excitation from the second excited level of the CH stretch mode
provides a two-electron process. However, this process can be discounted since the short
lifetime $\left(\gamma_{eh}^{(h)}\right)^{-1} \sim 1$~ps gives a nearly two orders of magnitude smaller rate.

Another possibility for the energy transfer is an anharmonic interaction of the CH stretch mode
with the reaction coordinate mode. In the simplest case this can be described  as a cubic coupling.
We expand the vibrational  Hamiltonian $H_{\mathrm{ph}}$
in Eq.~(\ref{Eq:Hfull}) up to cubic terms to account for the coupling between the directly
excited ($\nu = h$) and the reaction coordinate (RC)  ($\nu = r$) modes,
\begin{eqnarray}
\label{Eq:Hph}
H_{\mathrm{ph}} & = & H_0 + H_{\mathrm{ph},1} + H_{\mathrm{ph},2} \equiv
\sum_{\nu=h,r,i}\hbar\Omega_{\nu} b_{\nu}^{\dagger} b_{\nu} \\  \nonumber
&&
+ \mathcal{K}_{h,r,i}\left( b_r^{\dagger} b_i^{\dagger} b_{h} +\mathrm{h.c.}\right)
 +\frac{1}{2}\mathcal{K}_{h,r,r}\left( b_r^{\dagger} b_r^{\dagger} b_{h} +\mathrm{h.c.} \right)
,
\end{eqnarray}
where $\mathcal{K}_{h,r,i}, \mathcal{K}_{h,r,r}$ are the anharmonic coupling constants,
and $\nu =i$ is possibly some auxiliary (idler) vibrational mode excited simultaneously
with the RC mode.

\begin{figure}[b]
\begin{centering}
\includegraphics[width=0.98\columnwidth]{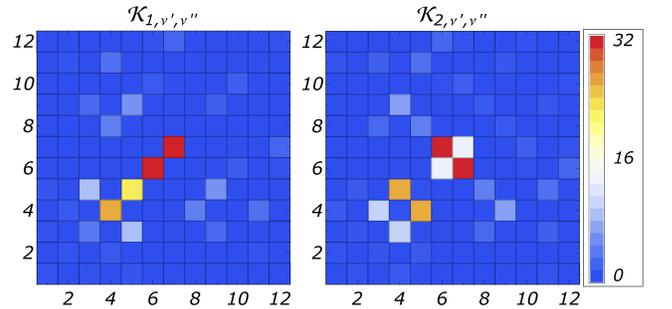}
\par
\end{centering}
\caption{Anharmonicity coupling constants $\mathcal{K}$ of the symmetric CH stretch mode \#1
(left panel) and asymmetric stretch mode \#2 (right panel) with other vibrational modes
of acetylene on Cu(001) surface. The color scheme is explained in the colorbar to the right.
Corresponding modes are given in table~\ref{tab:VibModes}.
Mode \#6 is the asymmetric rotation, and mode \#11 is the symmetric rotation.
\label{Fig:kappas}}
\end{figure}

DFT calculation of the anharmonic coupling constants is prohibitively time-consuming.
We thererfore construct (see in \textit{SM-II\&III}) a simpler model which takes into account
only the pair interactions between the nearest neighbors as springs on rods; the
results for the coupling coefficients are shown as colors in Fig~\ref{Fig:kappas}.
The symmetric CH stretch mode \#1 couples most efficiently with a pair of the asymmetric rotations \#6.
In contrast, asymmetric CH stretch mode \#2 couples strongly with a pair of non-equivalent phonons, i.e.,
asymmetric rotation \#6 and cartwheel mode \#7.
The coupling to the symmetric rotation mode \#11 is ineffective and the coefficients are approximately
two orders of magnitude smaller. Thus,  the asymmetric rotation  \#6 (CH out-of-plane bend) is the most
probable candidate for being the acetylene/Cu(001) rotation precursor.

\begin{figure}[t]
\begin{centering}
\includegraphics[width=0.5\columnwidth]{{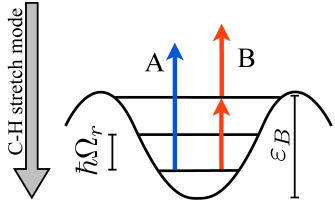}}
\par\end{centering}
\caption{\label{fig:ladderClimbing} Low-frequency mode ladder climbing process.}
\end{figure}

Two different pathways responsible for the two-electron partial process $R_B$ are possible.
If the barrier height is $\varepsilon_B > 2 \Omega_r$, only the second term in Eq.~(\ref{Eq:Hph}) contributes.
In this case the rotation starts from the excitation of the high energy symmetric CH stretch mode \#1 and
two successive decays into pairs of RC mode (\# 6) are needed to overcome the barrier. Such a overtone
ladder climbing process (see Fig.~\ref{fig:ladderClimbing}) was first discussed for a desorption of CO molecules
in Ref.~\onlinecite{Gao1994}.

If $\Omega_r<\varepsilon_B < 2 \Omega_r$, one pair of the RC stretch excitations is enough to cross the barrier which
just gives a correction to the single-electron rate $R_A$. In this case, the only possibility for a two-electron process comes from
the third term in Eq.~(\ref{Eq:Hph}). The rotation starts from the excitation of the high energy antisymmetric CH stretch mode (\#2),
and  two successive decays into pairs of RC mode (\# 6) and  idler mode (\#7) are needed to overcome the barrier.
This corresponds to the usual single-step ladder climbing process~\cite{Gao1997}.

In both cases the generation rate of RC phonons (pairs of phonons) is proportional to the
generation rate of the CH stretch phonons,
$
\Gamma_\mathrm{iet}(\Omega_r,V) \propto \Gamma_\mathrm{iet}(\Omega_h,V)
$ (see more detail in
\textit{SM-IV}).
Analysing the Pauli master equations for the one- and two-step  RC potential ladder climbing processes shows that the rotation rate is
\begin{equation}
\label{Eq:Rb}
R_B(V) = B \Gamma_{\mathrm{iet}}^{2}(\Omega_h,V),
\end{equation}
with different formulas for the coefficient $B$ depending on the corresponding anharmonic
coupling coefficients $\mathcal{K}$ and the ratio of the RC and idler modes linewidth to the
detuning $\Omega_h - \Omega_r - \Omega_i$ ($\Omega_h - 2\Omega_r$),
because the anharmonic decay of the stretch mode is nonresonant, see \textit{SM-V}.

The fitted value $B=6\times 10^{-16}\:\textrm{s}$ reproduce the experimental data above the higher
threshold $\sim 360$~mV well, see Fig.~\ref{fig:YvsV}. The hybridizations parameters
$\Delta_s=200$~meV, $\Delta_t=12$~meV were fixed from the tunneling current $I(\Omega_h)=40$~nA and
the fraction of the inelastic component $I_\mathrm{inel}/I\approx 1\%$ at $V=400$~mV. To simulate higher current levels,
the same parameters were used except $\Delta_t$ which changes as a function of bias and tunneling current as shown in Fig.~\ref{fig:YvsI}(b).
The calculated rotation rate as a function of tunneling current underlines the importance of the two-electron process at
high currents and voltages, see Fig.~\ref{fig:YvsI}(a). From the dash-dotted lines in Fig.~\ref{fig:YvsI}(a) which shows the
one and two-electron contributions where only the one-electron process,
excitation of the C-H mode followed by the anharmonic mode coupling to the RC mode,
is responsible for the reaction yield at low current below around $1$~nA.

The anharmonic coupling coefficients described above allows for a theoretical extimation of $B$.
If $\Omega_r<\varepsilon_B < 2 \Omega_r$, the estimate is $B=3.5\times10^{-15}$~s, which is
about 7 times larger than the best fit value. However, as noted above, the double excitation of the
reaction coordinate mode gives in this case a contribution to $R_A(V)$.
The resulting value of the coefficient $A$ is then $6\times 10^{-3}$, i.e.,  $10^3$ times larger than
the fitted value $2.5\times10^{-6}$. This allows us to exclude the case of a lower rotational barrier
from consideration since it would give rise to a much faster one-electron process than seen experimentally.
The only possibility left is that $\varepsilon_B > 2 \Omega_r$. The theoretical estimate for $B$ is then $\sim 2.4\times 10^{-15}$~s,
i.e., approximately 5 times larger than the fitted value. We believe, due to approximations, this value to be in reasonable agreement with
the experiment.

As can be seen from Fig.~\ref{fig:YvsV}, below the higher energy threshold $\Omega_h$ the rotation rate per electron
is very low but non zero above a lower threshold of $\sim$240~meV.
This can be attributed to an inelastic electron tunneling  processes which involves a
simultaneous combination band~\cite{Jakob1998}  generation of two coherent
phonons $\nu = c1,c2$. Assuming that the adsorbate energy in Eq.~(\ref{eq:He}) is now a function
of these vibrational modes, $\varepsilon_{a}(\{ q_{c1},q_{c2}\})$ , and expanding it in a Taylor series
\begin{equation}\label{eq:Hcomb}
\varepsilon_{a}(\{ q_{c1},q_{c2}\})=\varepsilon_{a}(\lbrace 0 \rbrace)+\eta(b_{c1}^{\dagger}+b_{c1})(b_{c2}^{\dagger}+b_{c,2}),
\end{equation}
where $\eta=\partial^{2}\varepsilon_{a}(\lbrace 0\rbrace)/\left(\partial q_{c1}\partial q_{c2}\right)$ and,
$b_{c1}$ and $b_{c2}$ are annihilation operators
of the vibrational modes with frequencies $\Omega_{c1}$ and $\Omega_{c2}$,  and $eh$ damping rates
$\gamma^{(c1)}_\mathrm{eh}$, $\gamma^{(c2)}_\mathrm{eh}$.
The analysis of this mechanism using the Keldysh-Green's functions
shows that the total coherent phonon excitation rate
takes  the form of a single-phonon excitation rate $\Gamma_{\mathrm{iet}}(\Omega,V)$~\cite{Tikhodeev2004},
Eq.~(\ref{Eq:Giet}), where a single vibrational frequency is simply replaced by  the sum of two vibrational frequencies.
This gives for the reaction rate, instead of Eq.~(\ref{Eq:Ra}),
\begin{equation}
\label{Eq:Rc}
R_C(V)=C \Gamma_\mathrm{iet}(\Omega_{c1}+\Omega_{c2},V),
\end{equation}
where $\Gamma_\mathrm{iet}(\Omega_{\mathrm{c},1}+\Omega_{\mathrm{c},2},V)=$
$$
\frac{\gamma^{(c1)}_\mathrm{eh}+\gamma^{(c2)}_\mathrm{eh}}{\Omega_{\mathrm{c},1}+\Omega_{\mathrm{c},2}}\frac{\Delta_t}{\Delta_{s}}
\left(\left|eV\right|-\Omega_{c1}-\Omega_{c2}\right) \Theta\left(\frac{\left|eV\right|}{\Omega_{c1}+\Omega_{c2}}-1\right)
$$
and the  coefficient $C$ depends on the parameters of the system as explained in \textit{SM-VI}.
Although it is difficult to make a theoretical estimate of this coefficient, the  fitted value to the experimental data is
 $C=1.2\times10^{-8}$. Thus, the combination band single-electron process is about $A/C \sim 200$ slower than the process
 with rotation excitation via CH stretch vibration. This is in reasonable agreement with the fact that the process Eq.~(\ref{eq:Hcomb})
 occurs in the next order of the perturbation theory  compared with the process Eq.~(\ref{eq:Ea}).

The over-barrier rotation mode excitation occurs due to an
inelastic tunneling generation of a coherent pair of phonons, e.g. the in-plane
bend or wag mode $\Omega_{c1} = 141$~meV and in-plane bend
or scissor mode $\Omega_{c2} = 118$~meV (modes \# 4 and 5). Another possibility
is that  the second of these phonons is the asymmetric out-of-plane rotation $\Omega_{c2} = 101$~meV
(mode \#6). The latter, being a hindered rotation mode,  may simplify the resulting molecule rotation.

To conclude,  we show that the precursor for the acetylene rotation on Cu(001) is the
out-of-plane bend (or asymmetric rotation) mode $\sim$100~meV (\# 6 in Table~\ref{tab:VibModes}).
Rotation with a higher threshold voltage 358 mV occurs if enough energy stored in
the high-frequency CH stretch mode excited by tunneling electrons is
transferred to the rotational mode (reaction coordinate). The
anharmonic coupling of the CH mode with the hindered rotation mode
is found to be responsible for a crossover from a single to multiple
electron process for tunneling currents higher than 10 nA. The lower
threshold voltage for rotation at 240 mV is attributed to a
combination band process of inelastic scattering of tunneling electrons on
a pair of lower-energy vibrational excitations of the acetylene molecule.

\begin{acknowledgments}
This work was supported in part by the Russian Ministry of Education and Science and the Russian Academy of Sciences.
H.U. was supported by a Grant-in-Aid for Scientific
Research (Grants No. S-21225001 and No. B-1834008) from
Japan Society for the Promotion of Science (JASP).
\end{acknowledgments}


%


\newpage
\clearpage

\begin{widetext}
\begin{center} {\large \bf Rotation of a single acetylene molecule on Cu(001) by tunneling electrons in STM: Supplementary Materials }

\bigskip 
Yulia E. Shchadilova,$^1$ Sergey G. Tikhodeev,$^{1,2,*}$ Magnus Paulsson,$^{3,2}$ and Hiromu Ueba$^{2}$\\
\medskip 
{\it \small
$^1$A. M. Prokhorov General Physical Institute, Russian Academy of Science, Moscow, Russia \\
$^2$Division of Nanotechnology and New Functional Material Science,\\
Graduate School of Science and Engineering, University of Toyama, Toyama, 930-8555 Japan \\
$^3$Department of Physics and Electrical Engineering, Linnaeus University, 391 82 Kalmar, Sweden\\
}
\end{center}



\setcounter{figure}{0}   \renewcommand{\thefigure}{S\arabic{figure}}

\setcounter{equation}{0} \renewcommand{\theequation}{S.\arabic{equation}}

\setcounter{section}{0} \renewcommand{\thesection}{S.\Roman{section}}

\renewcommand{\thesubsection}{S.\Roman{section}.\Alph{subsection}}

\makeatletter
\renewcommand*{\p@subsection}{}  
\makeatother

\renewcommand{\thesubsubsection}{S.\Roman{section}.\Alph{subsection}-\arabic{subsubsection}}

\makeatletter
\renewcommand*{\p@subsubsection}{}  
\makeatother



\begin{small}
The details are given for
\begin{center}
\begin{itemize}
\item DFT calculation of the vibrational modes of acetylene molecule on Cu(001) surface and the inelastic tunneling;
\item Third order anharmonic terms in the vibrational Hamiltonian which are responsible for the energy transfer between the vibrational modes;
\item Model calculation of the anharmonic coupling constants;
\item Anharmonic excitation rate of the hindered rotational phonons using Keldysh diagram technique;
\item Pauli master equation approach  to calculate the probability of rotations due to anharmonic decay of high frequency CH stretch mode;
\item Keldysh-Green's function  derivation of the combination band two-phonon rotation excitation.
\end{itemize}
\end{center}

{\footnotesize PACS numbers: 68.37.Ef, 68.43.Pq}
\end{small}
\end{widetext}

\section{DFT calculation results}\label{Appendix:DFT}

The DFT calculations were performed in SIESTA using a supercell of the Cu(001) with a 4x4 surface and 4 atoms thick slab.
Calculational details include a real space cutoff of 200 Ry, Gamma points approximation, double/single-z polarized (DZP/SZP) basis set for the C (DZP), H(DZP), and Cu(SZP) atoms.
Computational details for the vibrational frequencies, electron-phonon coupling, and electron-hole pair damping can be found in Ref.~\cite{Frederiksen2007}.
For the nudged elastic band (NEB) calculation more refined computational parameters were necessary, a 5x5 surface, 600 Ry cutoff, 3x3 k-points, and the DZP basis set for all species. Using the refined parameters changed vibrational frequencies by less than 10 \%.
Although the NEB calculation is fairly converged with computational parameters, the calculated barrier is still approximate
because of the basis set superposition error inherent in the SIESTA method. We therefore use the experimental barrier height in the main text.

\begin{figure}[h]
\begin{center}
\includegraphics[width=0.7\linewidth]{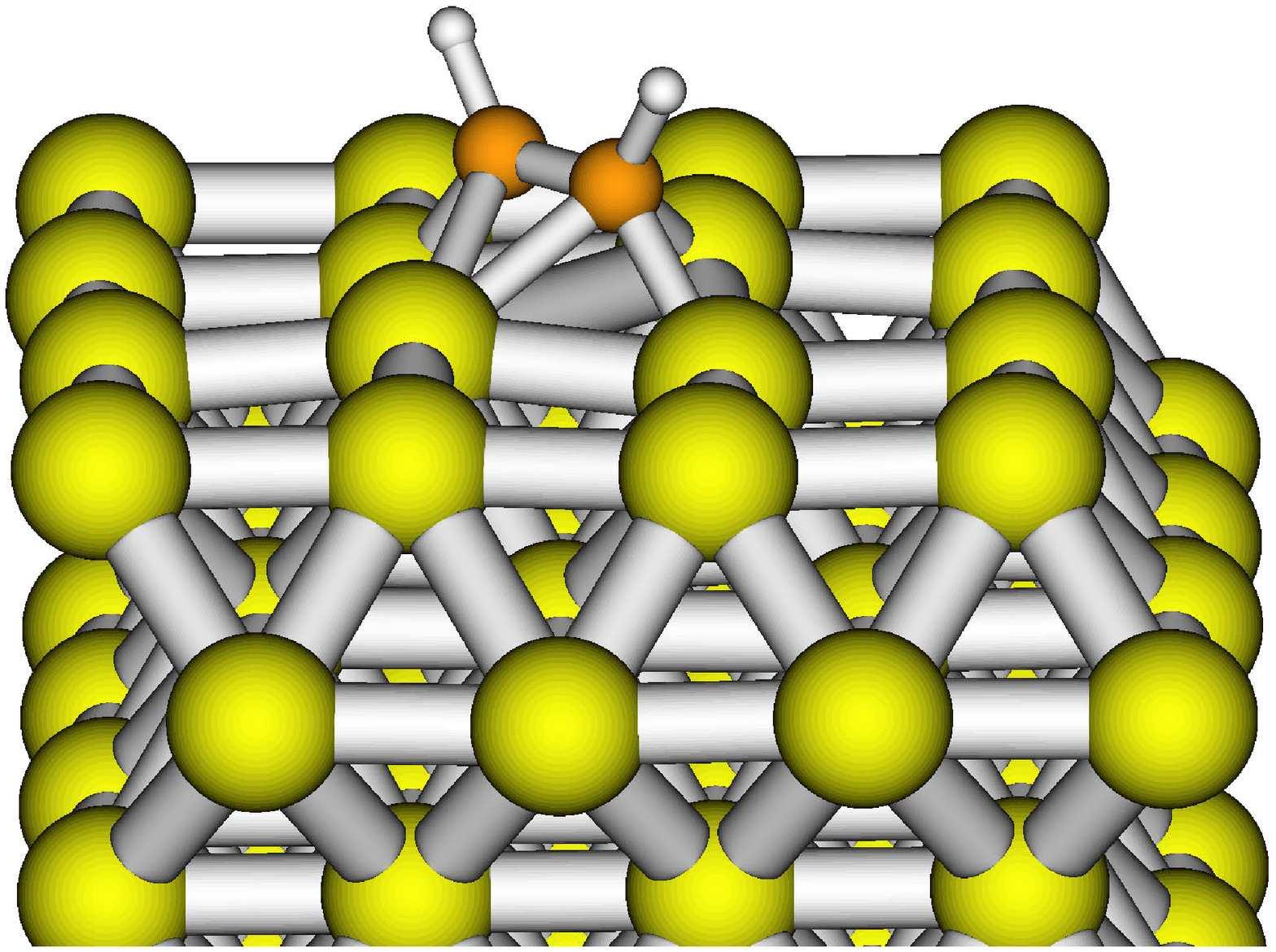}
\end{center}
\caption{\label{fig:C2H2config}
Calculated equilibrium configuration of C$_2$H$_2$ on Cu(001). The C-C and CH bond lengths and C-CH bond angle are given in the text.
}
\end{figure}

The  equilibrium configuration of C$_2$H$_2$ on Cu(001) and the relaxation of the Cu atoms is shown schematically in Fig.~\ref{fig:C2H2config}.
The results  for  the vibrational
energies $\hbar \Omega_\nu$ and  electron-hole damping rates $\gamma_{\mathrm{eh}}^{(\nu)}$, $\nu = 1, \ldots 12$  are shown in Tab.~I 
of the main text.

\begin{figure}[h]
\begin{tabular}{|c||c||c|}
\hline
1 \includegraphics[width=0.25\linewidth]{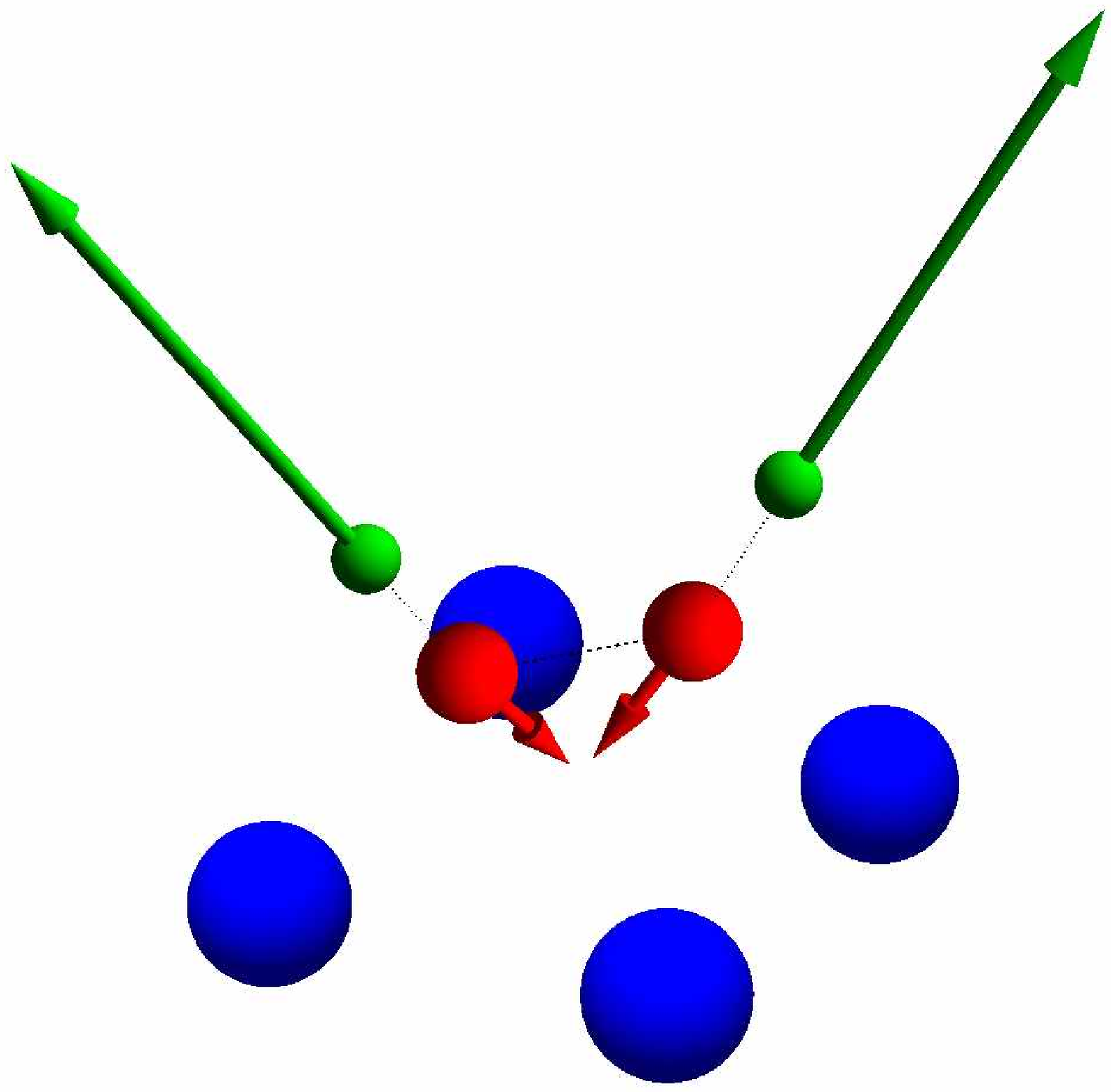} & 2 \includegraphics[width=0.25\linewidth]{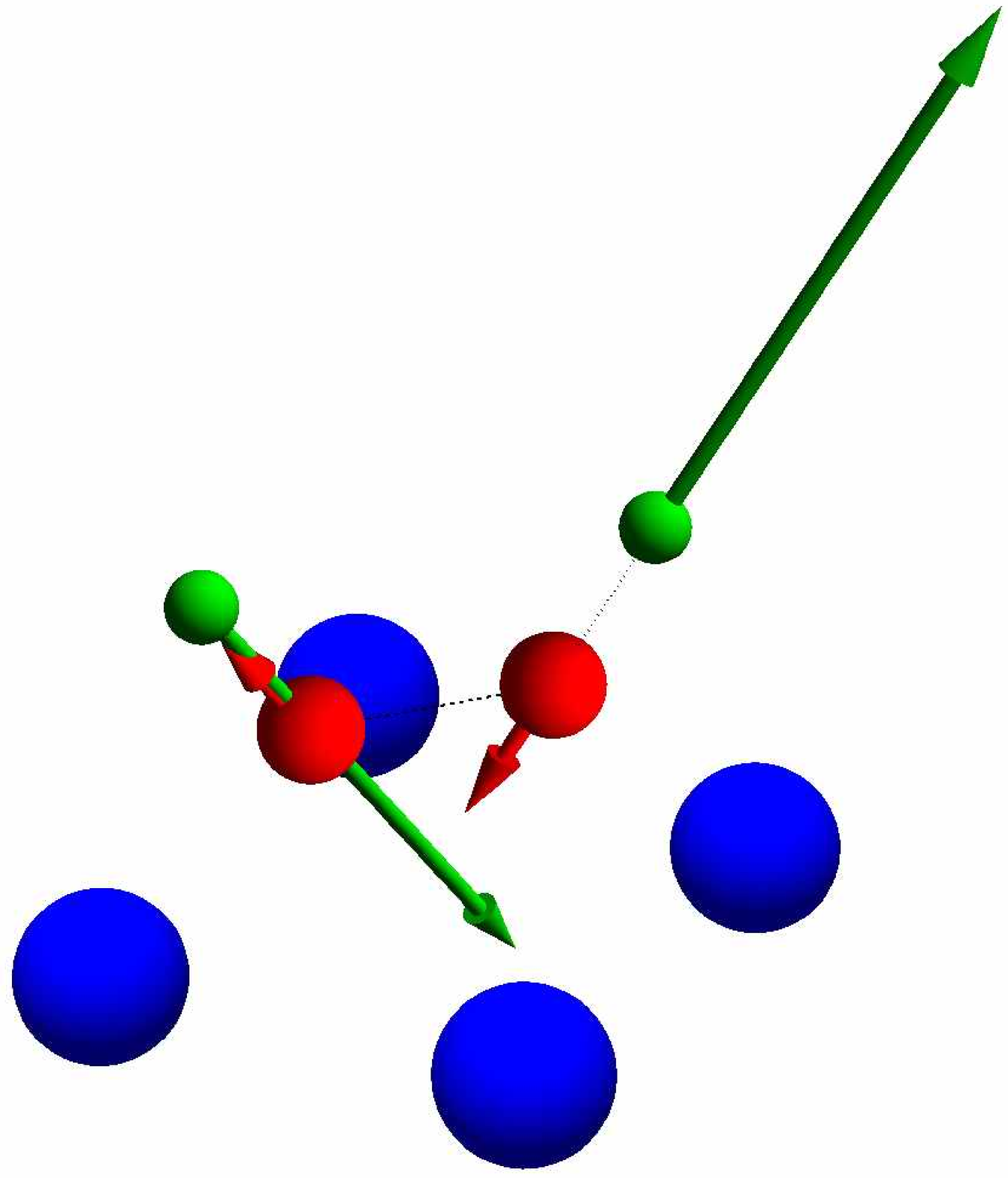} & 3 \includegraphics[width=0.25\linewidth]{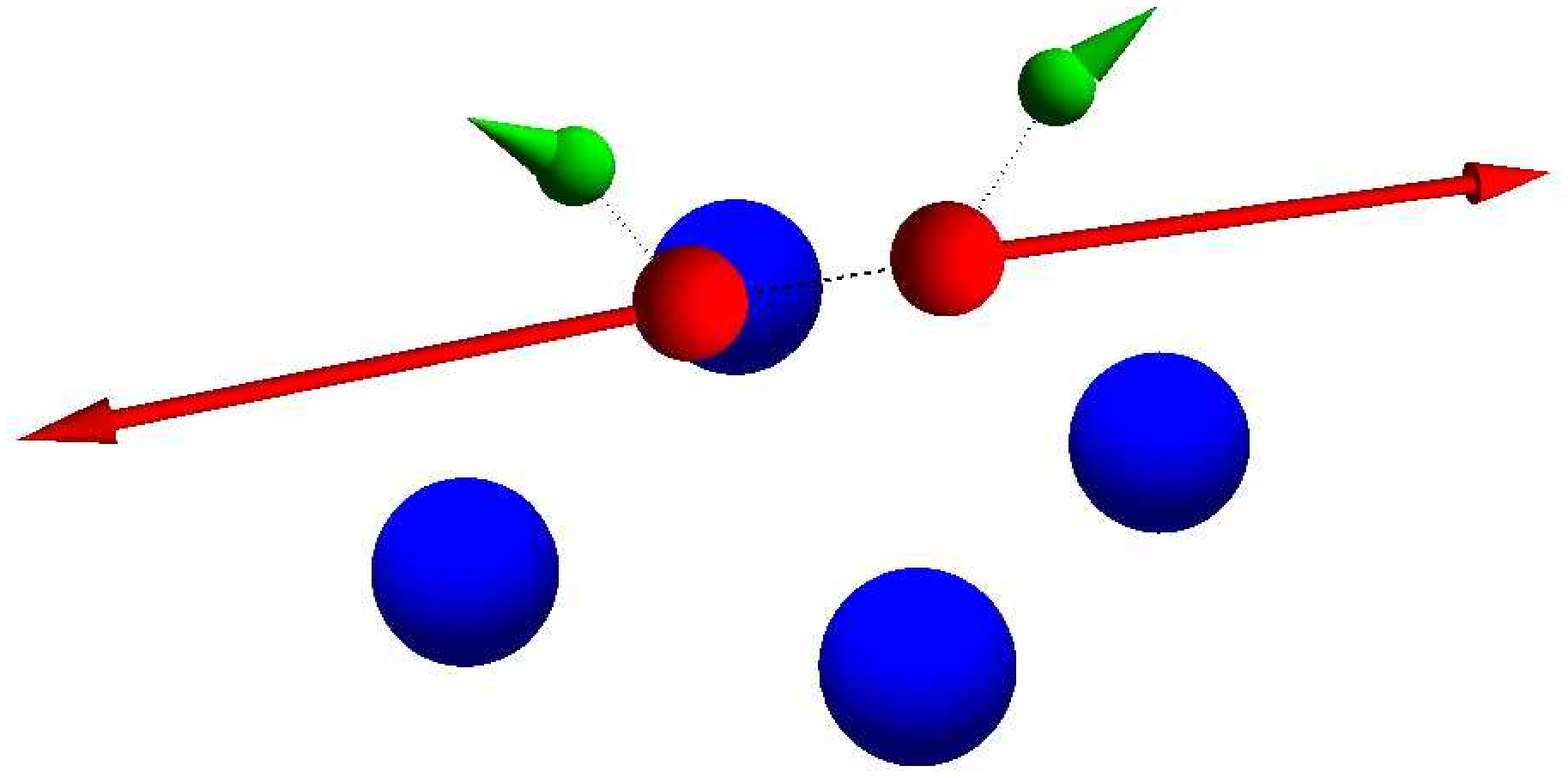}\tabularnewline
\hline
\hline
4\includegraphics[width=0.25\linewidth]{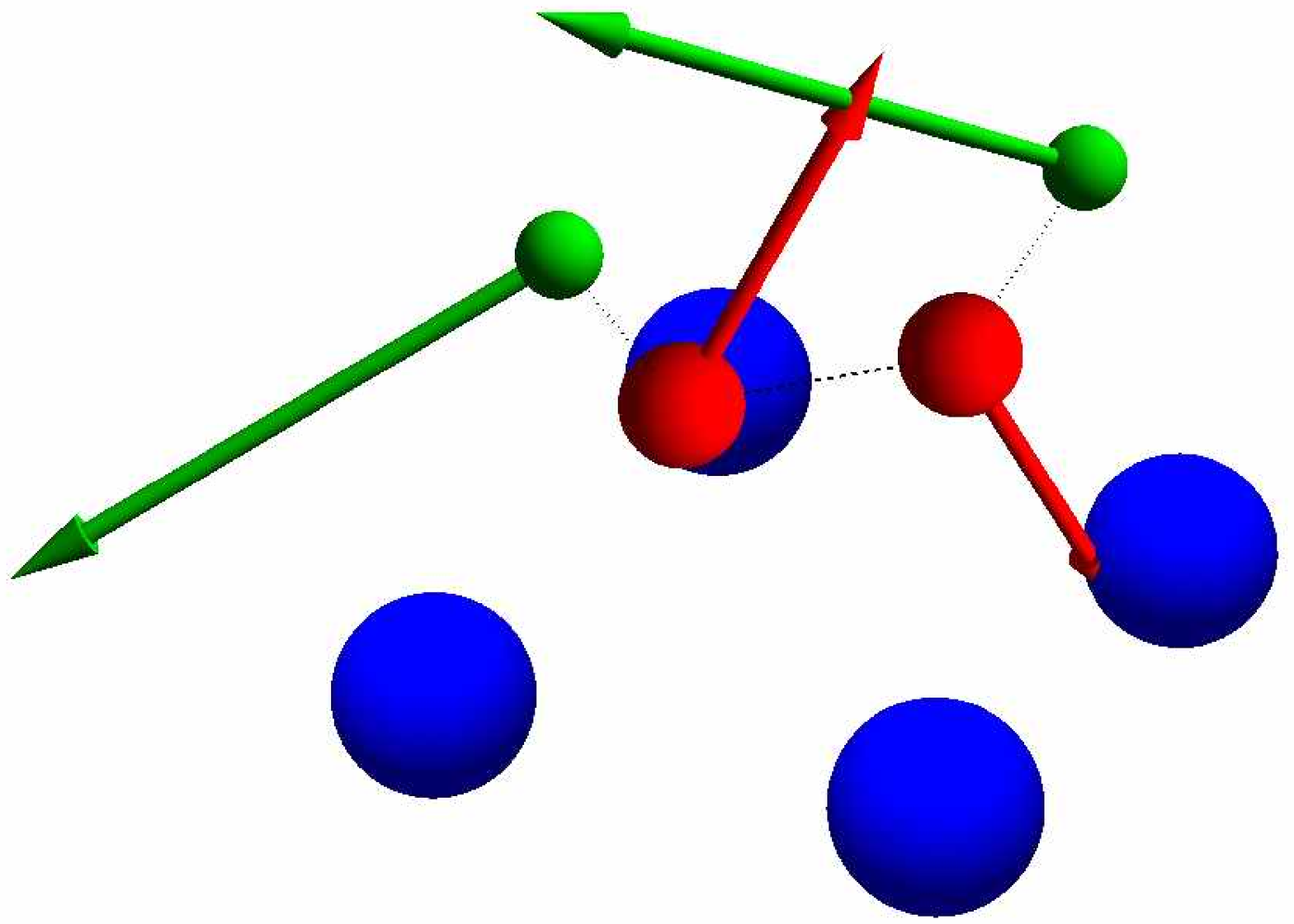} &5 \includegraphics[width=0.25\linewidth]{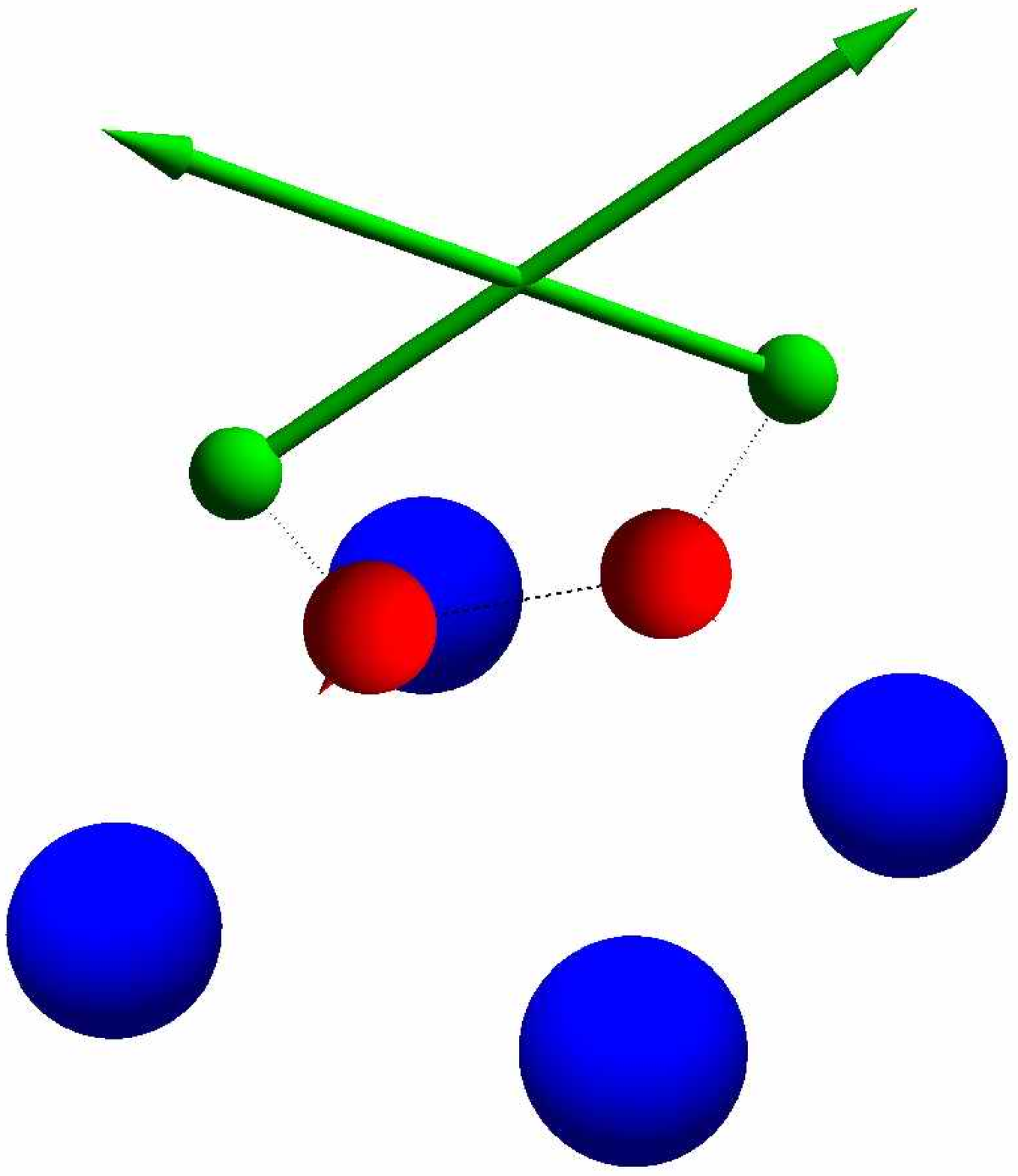} &6 \includegraphics[width=0.25\linewidth]{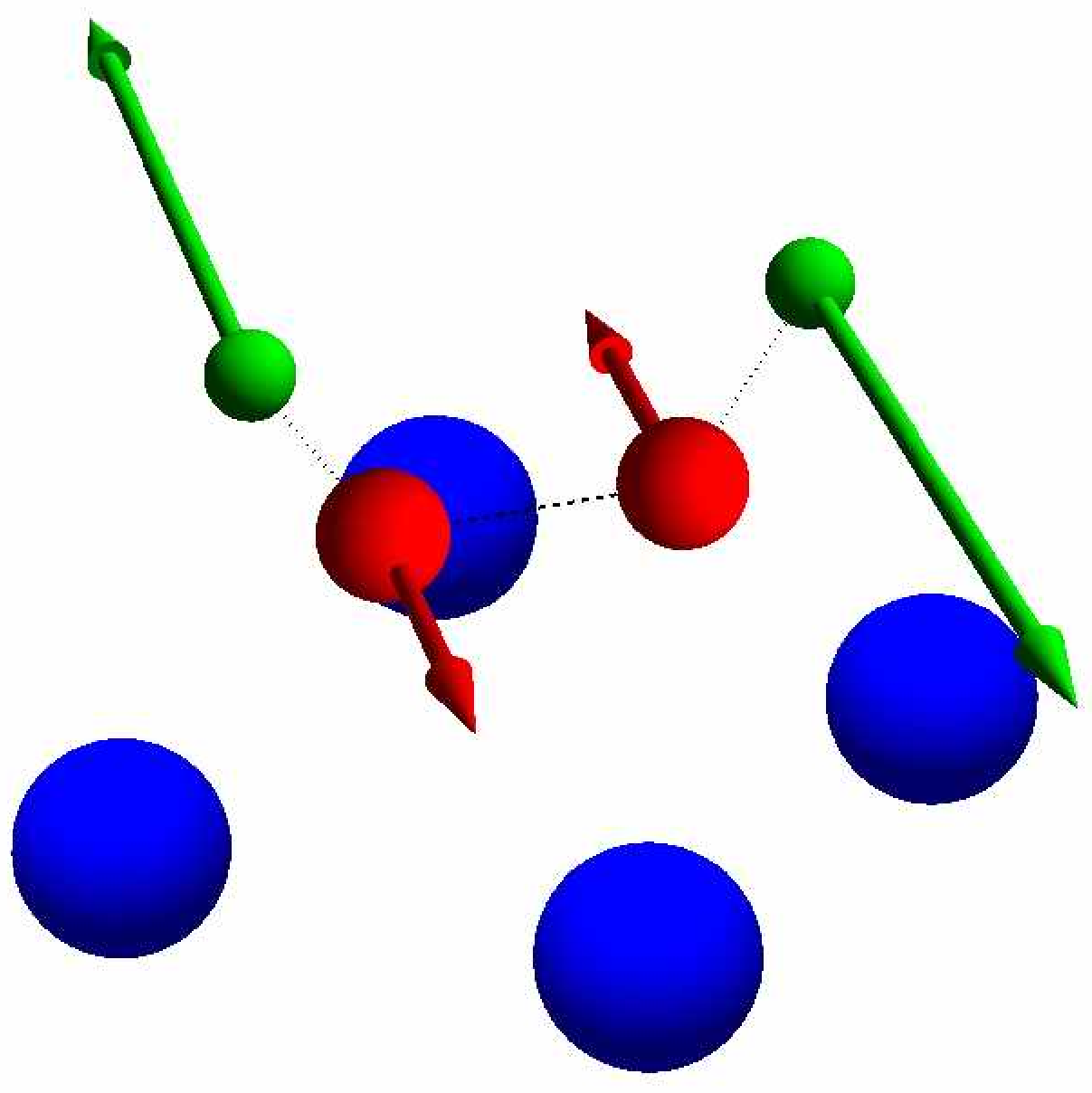}\tabularnewline
\hline
\hline
7\includegraphics[width=0.25\linewidth]{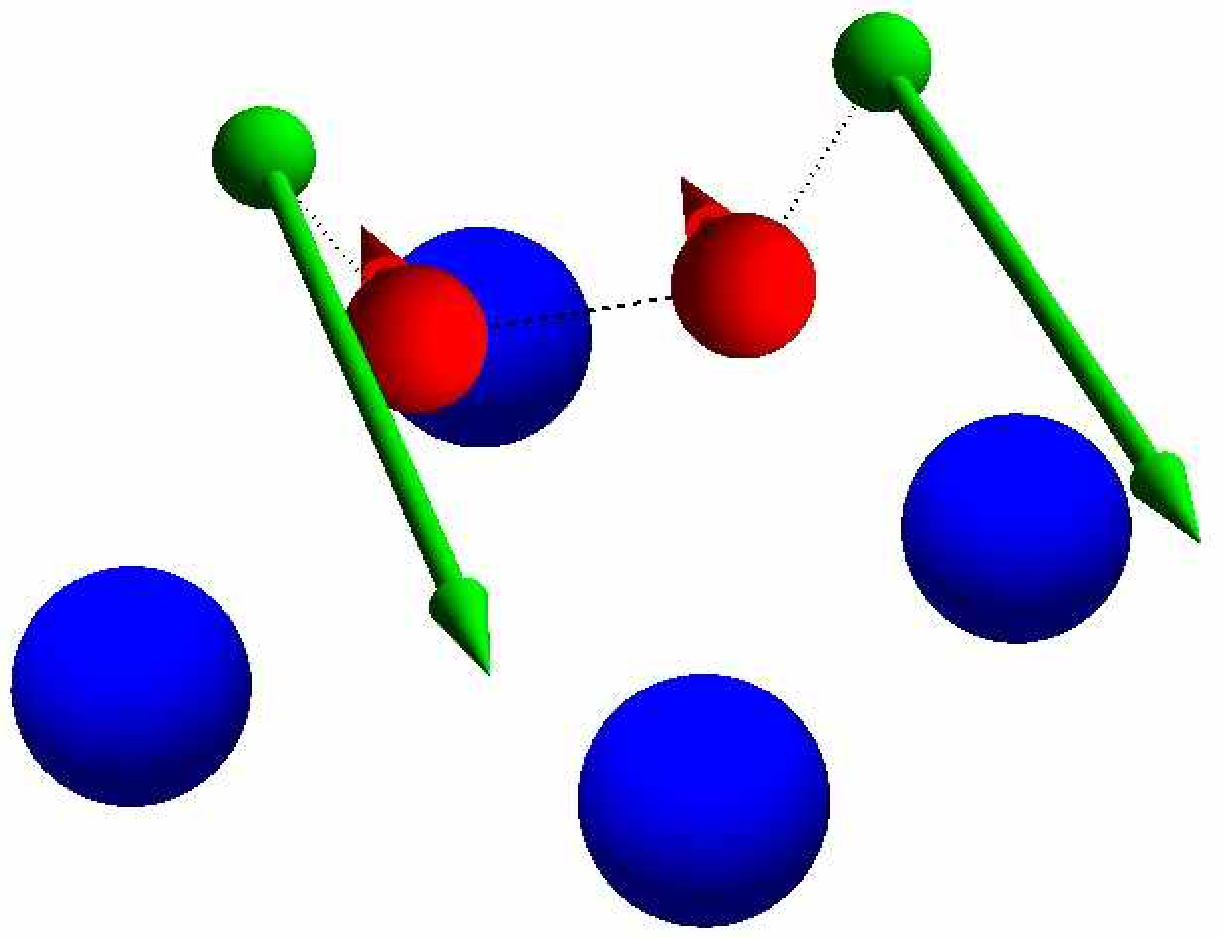} &8 \includegraphics[width=0.25\linewidth]{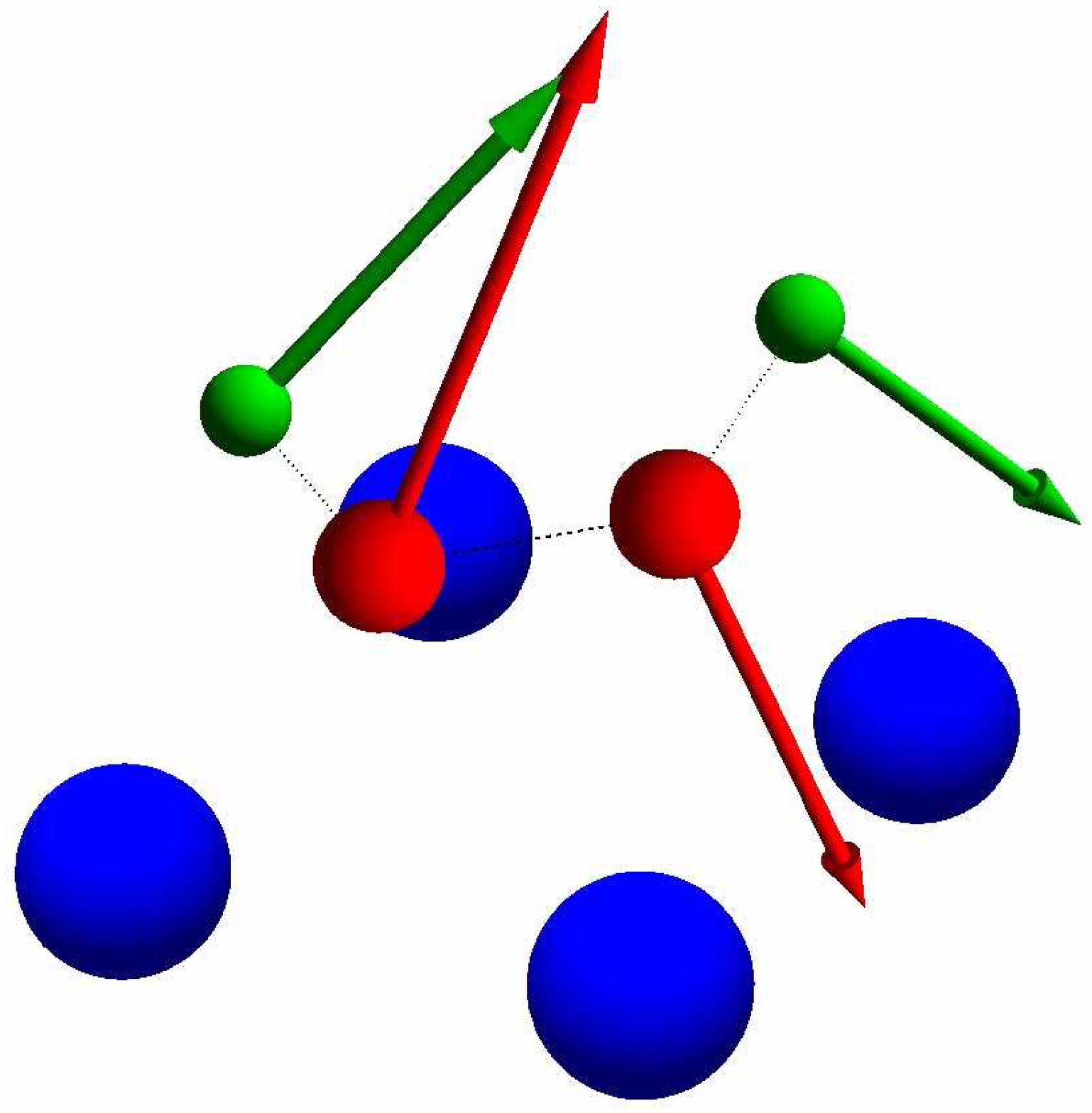} &9 \includegraphics[width=0.25\linewidth]{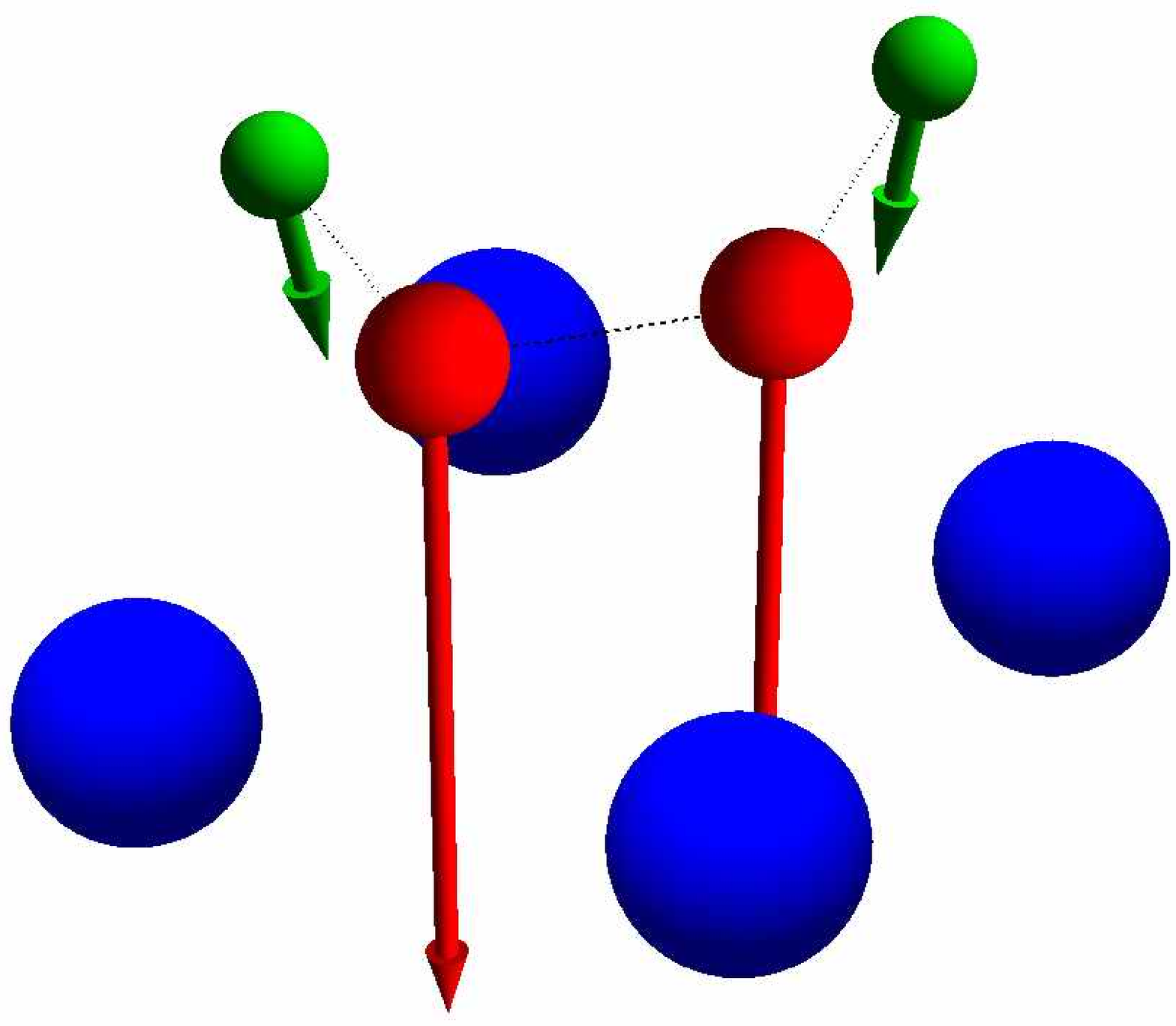}\tabularnewline
\hline
\hline
10 \includegraphics[width=0.25\linewidth]{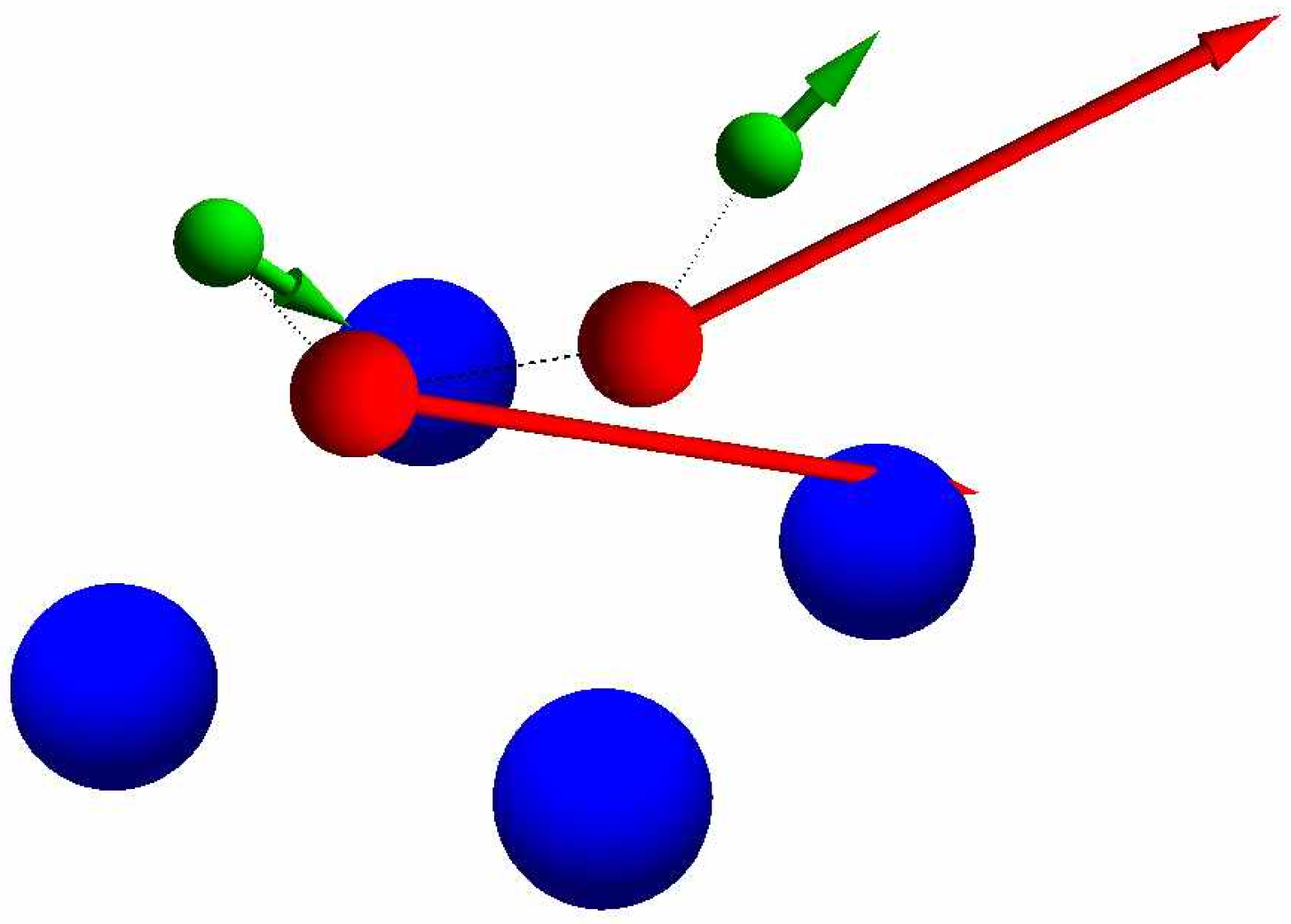} &11 \includegraphics[width=0.25\linewidth]{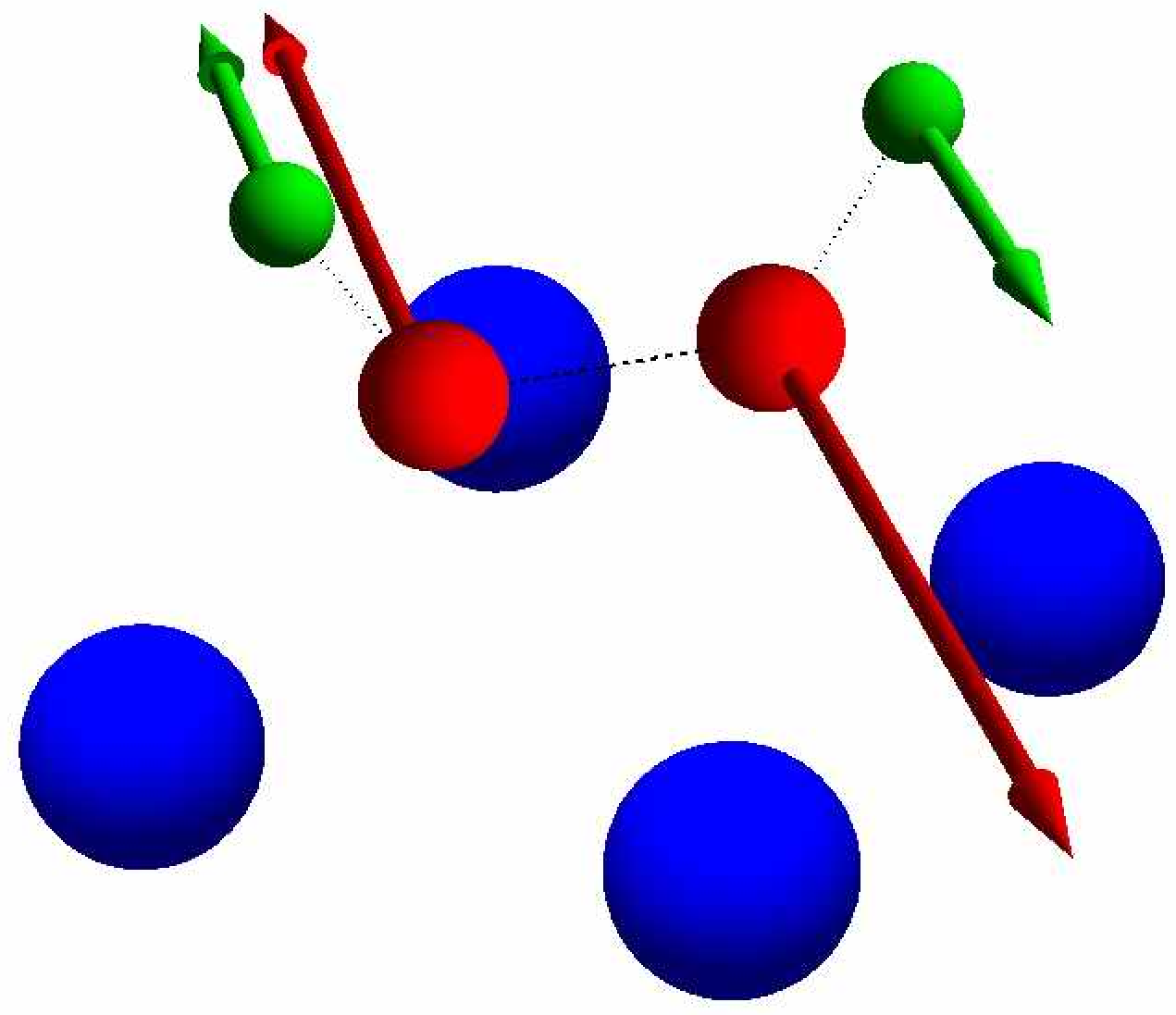} &12  \includegraphics[width=0.25\linewidth]{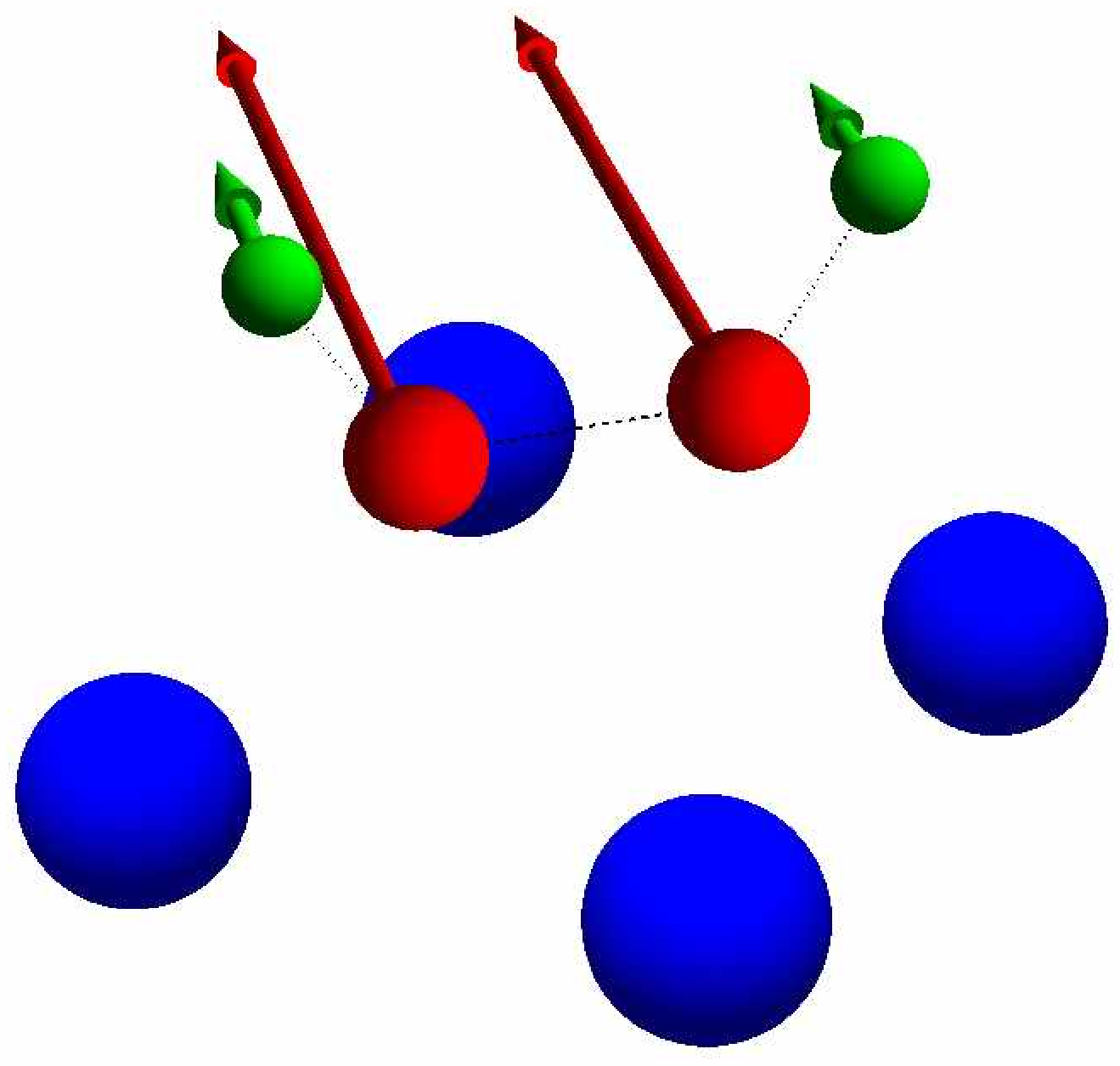}\tabularnewline
\hline
\end{tabular}
\caption{\label{fig:VibModes}
Vibrational modes of acetylene on Cu(001). Red  and green balls show the carbons and hydrogens of the adsorbed acetylene molecule,
blue balls are the four nearest neighbor copper atoms on the Cu(001) surface.}
\end{figure}

The calculated equilibrium configuration of the chemisorbed C$_2$H$_2$ molecule on Cu(001) is in agreement with
 Ref.~\onlinecite{Olsson2002}: $d_{\rm CC}=1.40$~\AA, $d_{\rm CH}=1.12$~\AA ~and bond angle CC-H is $120.7^\circ$.
The acetylene atom displacements
for all modes are illustrated in Fig.~\ref{fig:VibModes}, where only four Cu atoms nearest to the acetylene molecule are shown as blue balls.

\section{Anharmonic coupling the CH stretch modes with other vibrational modes of the adsorbed Acetylene}\label{Appendix:AnharmonicCoeff}

In this section we explicitly describe the scheme to calculate the normal coordinates modes of a system (in our case the system involves the  acetylene molecule and the surface of copper) and then the anharmonic couplings between them.

Shifting each atom from the equilibrium position $\lbrace \vec{R}_{i}\rbrace$  with a displacement vector $\lbrace\delta \vec{r}_{i}\rbrace$
and expanding the molecule potential energy $U(\lbrace\delta \vec{r}_{i}\rbrace)$ in Taylor series
\begin{equation}
U(\lbrace\delta \vec{r}_{i}\rbrace)=U_{0}+\frac{1}{2}\sum_{i,j}^{N}\sum_{\alpha,\beta=1}^{3}a_{ij}^{\alpha\beta}\delta r_{i}^{\alpha}\delta r_{j}^{\beta}+o\left(\left\Vert \delta r\right\Vert ^{2}\right),
\end{equation}
where $i,j$ denote the atomic indices,  $\alpha$, $\beta$  are the Cartesian indices,
$U_{0}$ is the potential energy of the molecule in the equilibrium state and coefficients
$a_{ij}^{\alpha\beta}$ are the matrix elements of the Hessian,
\begin{equation}
a_{ij}^{\alpha\beta}=\left.\frac{\partial^{2}U}{\partial\delta r_{i}^{\alpha}\partial\delta r_{j}^{\beta}}\right|_{r_{i}=R_{i}} .
\end{equation}
Equations of motion for shifted atoms can be written as
\begin{equation}\label{EqMotion}
m_{i}\frac{d^{2}}{dt^{2}}\delta r_{i}^{\alpha}=-\sum_{i,j}^{N}\sum_{\alpha,\beta=1}^{3}a_{ij}^{\alpha\beta}\delta r_{j}^{\beta} ,
\end{equation}
where $m_{i}$ is a mass of the $i$-th atom of the adsorbed molecule. For simplicity, we neglect the vibrational modes of the substrate,
assuming the copper atoms to be infinitely heavy, and fixed at relaxed equilibrium positions.
Substituting $\delta r_{i}^{\alpha}=\xi_{i}^{\alpha}\exp\left(-i\omega t\right)$ into Eqs.~(\ref{EqMotion})
one obtain $3N=12$ equations of motion
\begin{equation}
m_{i}\omega^{2}\xi_{i}^{\alpha}-\sum_{i,j}^{N}\sum_{\alpha,\beta=1}^{3}a_{ij}^{\alpha\beta}\xi_{j}^{\beta}=0.
\end{equation}
To find the normal modes of the system one needs to solve the secular equation
\begin{equation}\label{eq:secular}
\mathrm{Det}\left|m_{i}\omega^{2}\delta_{ij}\delta_{\alpha\beta}-a_{ij}^{\alpha\beta}\right|=0.
\end{equation}

In order to estimate the anharmonic couplings between different modes we expand the potential energy to the next order
\begin{multline}
U(\lbrace\delta\vec{r_{i}}\rbrace)=U_{0}+\frac{1}{2}\sum_{i,j}^{N}\sum_{\alpha,\beta=1}^{3}a_{ij}^{\alpha\beta}\delta r_{i}^{\alpha}\delta r_{j}^{\beta}+\\
+\frac{1}{6}\sum_{i,j,k}^{N}\sum_{\alpha,\beta,\gamma=1}^{3}b_{ijk}^{\alpha\beta\gamma}\delta r_{i}^{\alpha}\delta r_{j}^{\beta}\delta r_{k}^{\gamma}+o\left(\left\Vert \delta r\right\Vert ^{3}\right).
\end{multline}
After rotating  to the basis of normal coordinates we obtain
\begin{multline}
U(\lbrace\delta\vec{\tilde{q}}_{i}\rbrace)=U_{0}+\frac{1}{2}\sum_{m=1}^{3N}\omega_{m}^{2}\delta \tilde{q}_{m}\delta \tilde{q}_{m}+\\
+\frac{1}{6}\sum_{m,m',m''=1}^{3N}\mathcal{K}^{(c)}_{m,m',m''}\delta \tilde{q}_{m}\delta \tilde{q}_{m'}\delta \tilde{q}_{m''}+o\left(\left\Vert \delta \tilde{q}\right\Vert ^{3}\right).
\label{Eq:3rd}
\end{multline}
where $\lbrace  \delta\vec{\tilde{q}}_{i} \rbrace$ is a set of normal coordinates which represents a solution of the secular equation (\ref{eq:secular}) and $\mathcal{K}^{(c)}$ is classical anharmonic coupling between the vibrational modes of the systems.
The transformation  from the original Cartesian coordinates to the  normal coordinates basis has a form
\begin{equation}
\mathcal{K}_{m,m',m''}^{(c)}=\sum_{i,j,k}^{N}\sum_{\alpha,\beta,\gamma=1}^{3}
b_{ijk}^{\alpha\beta\gamma}\frac{e_{m,i}^{\alpha}}{\sqrt{m_i}}\frac{e_{m',j}^{\beta}}{\sqrt{m_j}}\frac{e_{m'',k}^{\gamma}}{\sqrt{m_k}},
\end{equation}
where $e_{m,i}^{\alpha}$ is the eigenvector coefficient between normal coordinate $\delta \tilde{q}_m$ and the shift of $i$-th atom in $\alpha$ direction $\delta r^{\alpha}_i$  - $\delta \tilde{q}_m = \sum_{i,\alpha} e^{\alpha}_{m,i}\delta r^{\alpha}_i$.

To quantize the vibrational modes, we introduce
a dimensionless displacement vector $\delta q_{m}$,
so that $\delta \tilde{q}_{m}=\delta q_{m}\sqrt{\hbar (2\omega_{m})^{-1}}$
and arrive from coordinates to operators in canonical way $\delta q_{m}\rightarrow \delta \hat{q}_{m}$. Then,
creation $b^{\dagger}_m$ and annihilation $b_m$ operators of the corresponding mode can be introduced
and the cubic term in Eq.~(\ref{Eq:3rd}) is transformed to
\begin{multline}\label{Eq:Anh}
H_{\mathrm{anh}}=\frac{1}{6}\sum_{m,m',m''=1}^{3N}\mathcal{K}_{m,m',m''}\\
\left(b_{m}^{\dagger}+b_{m}\right)\left(b_{m'}^{\dagger}+b_{m'}\right)\left(b_{m''}^{\dagger}+b_{m''}\right).
\end{multline}
where the anharmonic coefficient $\mathcal{K }_{m,m',m''}$ is connected with classical one $\mathcal{K }_{m,m',m''} = \mathcal{K}^{c}_{m,m',m''}\hbar^{3/2}\left(2\sqrt{2\omega_{m}\omega_{m'}\omega_{m''}}\right)^{-1}$.

\section{Springs on rods: a simpler model of acetylene on Cu(100)}\label{Appendix:SpringRod}
In order to estimate the anharmonic couplings $\mathcal{K}_{m,m',m''}$
we introduce the simpler model potential $U(\lbrace \vec{r}_i \rbrace)$ and imply the scheme discussed in Supplementary material~\ref{Appendix:AnharmonicCoeff}.

We propose
to describe the CH, CC and C-nearest neighbor Cu bonds by springs on rods (the latter
to fix the central character of the forces):
\begin{equation}
U_\mathrm{spr}(r_{i},r_{j})=\frac{\omega_{ij}^{2}}{2}\left(\sqrt{||r_{i}-r_{j}||^{2}}-L_{ij}\right)^{2},
\end{equation}
where $\omega_{ij}$ and $L_{ij}$ are the parameters (see in Tab.~\ref{tab:VibModes3}) chosen to reproduce the calculated by DFT set of eigenfrequencies.
The comparison of the DFT calculated eigenenergies with that within the simpler model are given in table~\ref{tab:VibModes2}.
\begingroup
\squeezetable
\begin{table}[h]
\caption{\label{tab:VibModes3}
The parameters used for the "springs on rods" model to reproduce the calculated by DFT vibrational modes}
\begin{ruledtabular}
\begin{tabular}{c c c c c}									
Bonds  &C-C 		&		C-H 		& C-Cu (nearest) 		&C-Cu (next-nearest)	      								   \\
$\omega$,~meV		& 159  & 370  & 70	& 25  \\
$L$, \AA  	& 1.39    & 1.04 & 1.94 & 2.23 \\
\end{tabular}
\end{ruledtabular}
\end{table}
\endgroup
\begingroup
\squeezetable
\begin{table}[h]
\caption{\label{tab:VibModes2}
Comparison between the vibrational eigenfrequencies $\hbar \Omega_\nu$ (in meV),
within the DFT
calculations (top row) and the simpler model.}
\begin{ruledtabular}
\begin{tabular}{c c c c c c c c c c c c c}
$\nu$ & 1            & 2            & 3           & 4           & 5        & 6          &  7       &  8        &  9       &  10    & 11     & 12 \\
DFT & 371        & 368        & 167       & 131       & 111    & 100      & 71      & 58       & 50      & 29     & 28     & 23  \\
model & 379.6        & 379.0      & 165.5   & 120.1   & 115.5   & 87.2   & 86.2   & 67.7    & 50.2    & 47.0  & 30.7  & 29.7\\
\end{tabular}
\end{ruledtabular}
\end{table}
\endgroup

We calculate the anharmonic coefficients $\mathcal{K}_{m,m',m''}$  and find the modes coupled effectively.
Figure~3 of the main text 
shows the dependence of coefficients $\mathcal{K}_{1,n,m}$ and $\mathcal{K}_{2,n,m}$
on mode numbers  $n$ and $m$. These coefficients indicate the most effective ways of decay
of the symmetric CH stretch mode \#1 and asymmetric CH stretch mode \#2.
These modes are known to be responsible for the high-energy threshold $\sim 360$ meV,
they are excited directly by the tunneling electrons.

The symmetric CH stretch mode \#1 decays most efficiently via excitation of a pair of equivalent phonons:
\#4 CH asymmetric in-plane bend or wag,
\#5 in-plane bend or scissor, \#6 out-of-plane bend or  asymmetric rotation, \#7 cartwheel.
Coupling of the symmetric CH stretch mode \#1 with the symmetric rotation mode \#11 is ineffective since the corresponding anharmonic
coefficient is $70$ times smaller than that for a coupling with the pair of out-of-plane bend or asymmetric rotation mode \#6.

The asymmetric CH stretch mode \#2 decays most efficiently via excitation of a pair of non-equivalent phonons, e.g.,
the pair of the asymmetric rotation \#6 and cartwheel mode \#7.

This simple estimation of the anharmonic coefficient shows that the rotation of the acetylene molecule
is initiated via excitation of the asymmetric rotation mode \#6.
There are two processes leading to the excitation of the reaction coordinate mode \#6, the excitation
of a pair of the asymmetric rotation phonons  or the excitation of one asymmetric rotation phonon and one
phonon of the cartwheel mode \#7.

\section{Double excitation of hindered-rotation mode}\label{Appendix:Double}
In this section we derive explicitly the excitation rate of the reaction coordinate mode.
As we have shown in Supplementary material~\ref{Appendix:SpringRod}, the excitation process
of the vibrational mode of the acetylene molecule  involves two possible pathways,
via excitation of the reaction coordinate mode and the auxiliary idler mode
or  via double excitation of the reaction coordinate mode
(terms $H_{\mathrm{ph},1}$ and $H_{\mathrm{ph},2}$  respectively,  in  Eq.~(8) of the main text). 

 We discuss here both of the scenarios using the Keldysh diagram technique.
In both cases under consideration the frequencies of the vibrational modes are far from the resonance,
$\Omega_{h} > 2 \Omega_{r} \sim \Omega_r+\Omega_i$ and the anharmonic interaction between them can be treated as weak.

In what follows we derive the excitation rate of the RC mode due to the process described by $H_{\mathrm{ph},1}$;
the excitation rate due to $H_{\mathrm{ph},2}$ can be obtained replacing the index $i$ of the idler
phonon  in all formulas below, with the index  $r$ of the RC phonon.

For the description of the effective stationary occupation densities of the RC
mode we use the kinetic equation.\cite{Tikhodeev2004}
The anharmonic component of the excitation rate of the RC mode is given by the
one-loop polarization operator. Neglecting the temperature corrections, it reads
\begin{multline}
\label{Eq:Ga}
\Gamma_{\mathrm{in}}(\omega,\Omega_r,V)= 2\pi \mathcal{K}_{h,r,i}^2\int n_{\mathrm{ph}}^{(h)}(\varepsilon+\omega)\rho_{\mathrm{ph}}^{(h)} (\varepsilon+\omega)\\
\left[ 1+ n_{\mathrm{ph}}^{(i)}(\varepsilon)\right] \rho_{\mathrm{ph}}^{(i)}(\varepsilon) d\varepsilon,
 \end{multline}
where $\rho_{\mathrm{ph}}^{(\nu)}(\varepsilon)$ is the density of states of the RC and high-frequency modes
and $n_{\mathrm{ph}}^{(\nu)}(\varepsilon)$ are the corresponding occupation densities.
Formula (\ref{Eq:Ga}) describes the energy transfer rate to the hindered rotation mode of an adsorbate due
to the anharmonic coupling with the CH stretch mode.

We proceed with the calculation of a total
excitation rate $\Gamma_{\mathrm{iet}}(\Omega_r,V)$ of the RC mode
\begin{equation}
\label{Gtot}
\Gamma_{\mathrm{iet}}(\Omega_r,V)=\int \Gamma_{\mathrm{in}}(\omega,\Omega_r,V)\rho_{\mathrm{ph}}^{(r)}(\omega) d \omega.
\end{equation}
After making a substitution of (\ref{Eq:Ga}) into (\ref{Gtot}) the total RC excitation rate takes the form
\begin{multline}\label{PhonGenRate}
\Gamma_{\mathrm{iet},2}(\Omega_r,V)\approx
2\pi \mathcal{K}^2_{h,r,i}  n_{\mathrm{ph}}^{(h)}(\Omega_i+\Omega_r) \rho_{\mathrm{ph}}^{(h)}(\Omega_i+\Omega_r) \\
+2\pi \mathcal{K}^2_{h,r,i}
n_{\mathrm{ph}}^{(h)}(\Omega_h) \left(\rho^{(r)}_\mathrm{ph}(\Omega_h-\Omega_i)+\rho_{\mathrm{ph}}^{(i)}(\Omega_h-\Omega_r)\right).
 \end{multline}
The second term in Eq.~(\ref{PhonGenRate}) shows a threshold dependence on  bias voltage,
because it is proportional to the high-frequency mode occupation numbers $n_{\mathrm{ph}}^{(h)}(\Omega_{h}) = \Gamma_{iet}(\Omega_{h})/2\gamma^{(h)}_{eh}(\Omega_{h})$.
It can be shown  that the first term in~(\ref{PhonGenRate}) can be omitted due to the fact that $n_{\mathrm{ph}}^{(h)}(\Omega_a+\Omega_r) \ll n_{\mathrm{ph}}^{(h)}(\Omega_{h})$.

The total excitation rate of the RC phonons due to the anharmonic term $H_{\mathrm{ph},1}$ is then
\begin{multline}\label{eq:PhonGenRate2}
 \Gamma_{\mathrm{iet},1}(\Omega_r,V)\approx  \\
 2\pi \mathcal{K}^2_{h,r,i}  \frac{\Gamma_{iet}(\Omega_{h})}{2\gamma^{(h)}_{eh}(\Omega_{h})}\left[\rho^{(r)}_\mathrm{ph}(\Omega_h-\Omega_i)+\rho_{\mathrm{ph}}^{(i)}(\Omega_h-\Omega_r)\right],
\end{multline}
and  due to  $H_{\mathrm{ph},2}$ is
\begin{equation}\label{eq:PhonGenRate3}
\Gamma_{\mathrm{iet},2}(\Omega_r,V)\approx 4\pi \mathcal{K}^2_{h,r,r} \frac{\Gamma_{iet}(\Omega_{h})}{2\gamma^{(h)}_{eh}(\Omega_{h})}\rho^{(r)}_\mathrm{ph}(\Omega_h-\Omega_r).
\end{equation}
Note that both rates are proportional to the CH stretch mode excitation rate $\Gamma_{iet}(\Omega_{h})$ and to small phonon densities of the RC and idler vibrational modes far from the
resonance.

\section{Ladder climbing with two- and one-step processes} \label{Appendix:Ladder}

In this section we calculate the excitation rate of the RC phonons $R_{B}(V)$  using Pauli master equations.
As  discussed in Supplement~\ref{Appendix:Double}, depending on the height of the rotational barrier,
two possible processes can lead to initiation of the rotations of the acetylene molecule on Cu(001).
In this section these processes will be considered separately as they involve  ladder climbing  of two different types.

In case when the rotational barrier height is $\Omega<\varepsilon_B<2\Omega$, the one-step ladder climbing process
takes place~\cite{Gao1994}. The excitation rate of this process $\Gamma_{\mathrm{iet},1}(\Omega_r,V)$ is given by (\ref{eq:PhonGenRate2})
and the relaxation rate is $\gamma_{\mathrm{eh}}^{(r)}$, thus the Pauli master equation can be written as
\begin{multline} \label{Eq:Pauli1step}
\frac{dP_{m}}{dt}  =  (m+1)\gamma_{\mathrm{eh}}^{(r)}P_{m+1}+m\Gamma_{\mathrm{iet},1}(\Omega_r,V)P_{m-1}\\
-  \left[m\gamma_{\mathrm{eh}}^{(r)}+(m+1)\Gamma_{\mathrm{iet},1}(\Omega_r,V)\right]P_{m}.
\end{multline}
The stationary solutions (in respect to $P_{0}$) for $m=1$ states in the localization potential of the RC mode
can be written as $P_{0}=1$, $P_{1}=\Gamma_{\mathrm{iet},1}(\Omega_r,V)\left(\gamma_{\mathrm{eh}}^{(r)}\right)^{-1}P_{0}\ll P_{0}$.

Reaction rate $R_B(V)$ is defined as a probability rate to overcome the localization potential barrier and in our notations
it is the excitation rate from the first excited level
\begin{equation}\label{eq:RB1}
R^{(1)}_B(V)=2\Gamma_{\mathrm{iet},1}^2(\Omega_r,V)\left(\gamma_{\mathrm{eh}}^{(r)}\right)^{-1}.
\end{equation}

If the rotational barrier height is $2\Omega<\varepsilon_B<3\Omega$ then  the excitation rate is assumed to be dominated by
the two-phonon anharmonic coupling with the high-frequency mode. The pair of RC phonons excitation rate $\Gamma_{\mathrm{iet},2}(\Omega_r,V)$ is given by (\ref{eq:PhonGenRate3})
and the de-excitation process is dominated by the single phonon relaxation rate $\gamma_{\mathrm{eh}}^{(r)}$. Then, according to Ref.~\onlinecite{Gao1994}, the Pauli master equation takes form
\begin{multline}\label{Eq:Pauli2step}
\frac{dP_{m}}{dt}=(m+1)\gamma_{\mathrm{eh}}^{(r)}P_{m+1} +m(m-1)\Gamma_{\mathrm{iet},2}(\Omega_r,V)P_{m-2}\\
-\left[m\gamma_{\mathrm{eh}}^{(r)}+(m+2)(m+1)\Gamma_{\mathrm{iet},2}(\Omega_r,V)\right]P_{m}.
\end{multline}
The stationary solutions (in respect to $P_{0}$) for $m=0,1,2$ states in the localization potential of the RC mode
can be written as $P_{0}=1$, $P_{1}=2\Gamma_{\mathrm{iet},2}(\Omega_r,V)\left(\gamma_{\mathrm{eh}}^{(r)}\right)^{-1}P_{0}\ll P_{0}$ and
\begin{equation}
P_{2}=\frac{\left(\gamma_{\mathrm{eh}}^{(r)} + 6\Gamma_{\mathrm{iet},2}(\Omega_r,V)\right)}{2\gamma_{\mathrm{eh}}^{(r)}}P_{1}\approx\frac{\Gamma_{\mathrm{iet},2}(\Omega_r,V)}{\gamma_{\mathrm{eh}}^{(r)}}P_{0}.
\end{equation}

Reaction rate $R_B(V)$ in this case is a sum of the excitation rates from the first excited state $6\Gamma_{\mathrm{iet},2}(\Omega_r,V)P_1$ and from  the second excited state $12\Gamma_{\mathrm{iet},2}(\Omega_r,V)P_{2}$,
\begin{equation}\label{eq:RB2}
R^{(2)}_B(V)=24\Gamma_{\mathrm{iet},2}^2(\Omega_r,V)\left(\gamma_{\mathrm{eh}}^{(r)}\right)^{-1}.
\end{equation}

The reaction rate in both cases is a quadratic function of the RC phonon excitation rate which is a feature of the two-step ladder climbing process
and differs only in a proportionality coefficient. For a more accurate estimation of proportionality coefficient
between the second power inelastic tunneling current and rotation rate
we need to take into account the fact that the inelastic tunneling current is a sum of several components which
arise from the scattering of tunneling elections on all relevant vibrational modes.
In our case we are interested in two CH stretch modes, symmetric \#1 and asymmetric \#2.
According to our DFT calculation the inelastic transmission through these modes are
$T^{(1)}=1.3\times10^{11}$~(s$\times$V)$^{-1}$ and $T^{(2)}=5.9\times10^{11}$~(s$\times$V)$^{-1}$.
In what follows we will take this into account. introducing the  probability factors to excite the symmetric CH
stretch mode $\zeta=T^{(1)}\left(T^{(1)}+T^{(2)}\right)^{-1}=0.18$ and the asymmetric one $(1-\zeta)=0.82$.

Substituting the expression for the excitation rate of the RC phonons (\ref{eq:PhonGenRate2}) and (\ref{eq:PhonGenRate3}) into
Eqs.~(\ref{eq:RB1}) and ~(\ref{eq:RB2}) correspondingly and using the expression
for the high-frequency phonons
$n_{\mathrm{ph}}^{(h)}(\Omega_{h})=\Gamma_{\mathrm{iet}}(\Omega_{h},V)\left(2\gamma_\mathrm{eh}^{(h)}\right)^{-1}$
occupation densities
we obtain the proportionality coefficient $B^{(1)}$ [Eq.~(9) of the main text] 
between the excitation reaction rate  and the phonon generation rate
 $R^{(1)}_{B}(V)=B^{(1)}\Gamma_{\mathrm{iet},1}(\Omega_{h},V)$.
\begin{multline}
B^{(1)}=2\pi^2 \mathcal{K}^4_{h,r,a} (1-\zeta)^2\frac{1}{\left(\gamma_\mathrm{eh}^{(h)}\right)^{2} \gamma_{\mathrm{eh}}^{(r)}}\\
\left(\rho^{(r)}_\mathrm{ph}(\Omega_h-\Omega_i)+\rho_{\mathrm{ph}}^{(i)}(\Omega_h-\Omega_r)\right)^2.
\end{multline}
Using $\mathcal{K}_{h,r,i}\approx 32$~meV, $\gamma^{(a)} = 0.2$~ps$^{-1}$, $\gamma^{(h)}=0.7$~ps$^{-1}$, $\gamma_r=0.7$~ps$^{-1}$, $\Delta=\Omega_h-\Omega_r-\Omega_i\approx 208$~meV ($\Omega_i\approx\Omega_r$)
we obtain $B^{(1)}=3.5\times10^{-15}$ ~s. It is $7$ times larger than  the coefficient $B=6\times 10^{-16}$~s obtained from the best fit of the experimental data.

Moreover, in  case of a  lower reaction barrier $\Omega<\varepsilon_B<2\Omega$ the process of double excitation of
the reaction coordinate phonons  gives a contribution to the linear part $R_A(V)$ of the rotation probability.
The rotation rate can be estimated then as $R_A(V)\approx \Gamma_{\mathrm{iet},2}$.
Then the impact of this process into the proportionality coefficient $A$ can be written as
\begin{equation}
A^{(2)}= 4\pi \mathcal{K}^2_{h,r,r} \zeta \tau^{(h)} \rho^{(r)}_\mathrm{ph}(\Omega_h-\Omega_r).
\end{equation}
Using the same parameters as above for the coefficient $B^{(1)}$, we arrive to the estimated value of $A^{(2)}=6\times 10^{-3}$,
which is three orders of magnitude larger than the value of $A$ obtained from the fitting of the experimental data.
Evidently, this makes the case of a lower reaction coordinate
barrier $\Omega<\varepsilon_B<2\Omega$ hardly possible .

Analogously, the coefficient $B^{(2)}$ between $R^{(2)}_{B}(V)$ and $\Gamma_{\mathrm{iet},2}^2(\Omega_{h},V)$,
\begin{equation}\label{eq:RB3}
B^{(2)}=48\pi^2 \mathcal{K}^4_{h,r,r} \zeta^2 \frac{\left(\rho^{(r)}_\mathrm{ph}(\Omega_h-\Omega_r)\right)^2}{\left(\gamma_\mathrm{eh}^{(h)}\right)^{2}\gamma_{\mathrm{eh}}^{(r)}}.
\end{equation}
Using $\mathcal{K}_{h,r,r}\approx 32$~meV, $\gamma^{(h)}_\mathrm{ph}=\gamma^{(1)}=1$~ps$^{-1}$, $\gamma^{(r)}=0.7$~ps$^{-1}$, $\Delta=\Omega_h-2\Omega_r\approx 208$ meV  we
obtain that $B^{(2)}\approx 2.4\times 10^{-15}$~s, $5$ times larger than the best fit value $B=6\times 10^{-16}$~s.

Thus the estimated value of the coefficient $B$ is slightly larger than the value obtained from the fit of the experimental data.
We believe that our simpler model overestimates the anharmonic coupling coefficients $\mathcal{K}_{h,r,r}$; for a better agreement
with the  experimental data we have to take $\approx 1.5$ times smaller values  $\mathcal{K}_{h,r,r}\sim 20$~meV.

\section{Combination band single electron process: Keldysh technique formulation} \label{Appendix:Combinational}
In this section we show  that  in the limit of a low temperature $T=0$  the excitation rate of  phonons excited simultaneously
via tunneling electron scattering can be written in the same form as  for the single-electron excitation.~\cite{Tikhodeev2004}

In order to calculate an excitation rate of coherent phonons we use the Keldysh-Green's
function method.\cite{Keldysh1965,Tikhodeev2004}
The kinetic equation for phonons
takes the form
\begin{equation}
\frac{\partial N_{\mathrm{c},i}}{\partial t}  =  \int\left[\Pi_{\mathrm{c},i}^{\mathtt{+-}}\left(\omega\right)D_{\mathrm{c},i}^{-+}(\omega)-
\Pi_{\mathrm{c},i}^{-+}(\omega)D_{\mathrm{c},i}^{+-}(\omega)\right]\frac{d\omega}{2\pi},
\end{equation}
where $i=1,2$; $N_{\mathrm{c},i}$ is a total number of phonons in corresponding mode. $D_{\mathrm{c},i}^{\pm\mp}$
are phonon Keldysh-Green's functions and $\Pi_{\mathrm{c},i}^{\pm\mp}$ are phonon polarization
operators. The former are given by
\begin{eqnarray}
D_{\mathrm{c},i}^{-+}(\omega) & = & -2i\pi\left(1+n_{\mathrm{ph}}^{({\mathrm{c},i})}(\omega)\right)\rho_{\mathrm{ph}}^{({\mathrm{c},i})}(\omega),\\
D_{\mathrm{c},i}^{+-}(\omega) & = & -2i\pi n_{\mathrm{ph}}^{({\mathrm{c},i})}(\omega)\rho_{\mathrm{ph}}^{({\mathrm{c},i})}(\omega),
\end{eqnarray}
where $n_{\mathrm{ph}}^{({\mathrm{c},i})}(\omega)$ is the vibration occupation function of $i$-th mode to be
determined self-consistently from the kinetic equation.

Substituting $D_{\mathrm{c},i}^{\pm\mp}(\omega)$ in the kinetic equation, we
obtain
\begin{multline}
\frac{\partial N_{\mathrm{c},i}}{\partial t} =  (-i)\int d\omega\rho_{\mathrm{ph}}^{({\mathrm{c},i})}(\omega)\left[\Pi_{\mathrm{c},i}^{+-}(\omega)-\right.\\
\left.-\left(\Pi_{\mathrm{c},i}^{-+}(\omega)-\Pi_{\mathrm{c},i}^{+-}(\omega)\right)n_{\mathrm{ph}}^{({\mathrm{c},i})}(\omega)\right].
\end{multline}
At low temperatures, $k_{B}T\ll\Omega_{\mathrm{c},i}$
the occupation density takes the form $n_{\mathrm{ph}}^{(\mathrm{c},i)}(\omega)\approx \Gamma_{\mathrm{in}}(\omega,\Omega_{\mathrm{c},i},V)\left(\gamma_{\mathrm{tot}}^{(\mathrm{c},i)}(\omega)\right)^{-1}$,
where $\Gamma_{\mathrm{in}}(\omega,\Omega_{\mathrm{c},i},V)$ and $\gamma_{\mathrm{tot}}^{(\mathrm{c},i)}$ are the excitation and relaxation
rates of the corresponding phonon modes. Then $\Gamma_{\mathrm{in}}(\omega,\Omega_{\mathrm{c},i},V)=\Pi_{\mathrm{c},i}^{+-}(\omega)$.

The self-energy $\Pi_{\mathrm{c},1}^{+-}(\omega)$, expanded to the second order in $\eta$ [see Eq.~(10) of the main text] 
reads
\begin{eqnarray}
\Gamma_{\mathrm{in}}(\omega,\Omega_{\mathrm{c},1},V)& =& \Pi_{\mathrm{c},1}^{+-}(\omega) =  -i\eta^{2}\int\frac{d\omega'}{2\pi}\frac{d\varepsilon}{2\pi}\\
 &  & D_{\mathrm{c},2}^{-+}(\omega')G_{a}^{+-}(\varepsilon)G_{a}^{-+}(\varepsilon-\omega-\omega'),\nonumber
\end{eqnarray}
where $G_{a}(\varepsilon)$ are the Keldysh-Green's functions of the electrons of the adsorbate.
Substituting $D_{\mathrm{c},2}^{-+}(\omega')$ and taking into account that $n_{\mathrm{ph}}^{(\mathrm{c},i)}(\omega)\ll1$,
we arrive to the following expressions for the polarization operators
\begin{equation}
\Gamma_{\mathrm{in}}(\omega,\Omega_{\mathrm{c},1},V)=\eta^{2}\int\frac{d\varepsilon}{2\pi}G_{a}^{+-}(\varepsilon)G_{a}^{-+}(\varepsilon-\omega-\Omega_{\mathrm{c},2}).
\end{equation}
In full analogy we obtain the expression for $\Gamma_{\mathrm{in}}(\omega,\Omega_{\mathrm{c},2},V)$.
Total coherent phonon excitation rate $\Gamma_{coh}=\Gamma_\mathrm{iet}(\Omega_{\mathrm{c},1},V)=\Gamma_\mathrm{iet}(\Omega_{\mathrm{c},2},V)\equiv \int \Gamma_{\mathrm{in}}(\omega,\Omega_{\mathrm{c},i},V)\rho_{\mathrm{ph}}^{(\mathrm{c},i)}(\omega)d\omega$ takes the same form for both modes
\begin{equation}
\Gamma_{coh}=2\pi\eta^{2}\int G_{a}^{-+}(\varepsilon-\Omega_{\mathrm{c},1}-\Omega_{\mathrm{c},2})G_{a}^{+-}(\varepsilon)d\varepsilon.
\end{equation}
The expression in a limit $T=0$ and $\Delta_{t}\ll\Delta_{s}$ can be approximated as
\begin{multline}\label{eq:G_coh}
\Gamma_\mathrm{coh}(V)\approx \\\frac{\gamma_\mathrm{coh}\left(\Omega_{\mathrm{c},1}+\Omega_{\mathrm{c},2}\right)}{\Omega_{\mathrm{c},1}+\Omega_{\mathrm{c},2}}\frac{\Delta_t}{\Delta_{s}}
F(\Omega_{\mathrm{c},1}+\Omega_{\mathrm{c},2},\;eV),
\end{multline}
where $F(\Omega,\;eV)=\left( eV-\Omega\right)
\theta \left(|eV|-\Omega\right)$
 and $\gamma_\mathrm{coh}(\omega)=2\pi \eta^2 \rho_a^2(\varepsilon_F)\omega$ is the relaxation rate
of the coherent phonons. We can use the inverse lifetime of the coherent modes
$\gamma^{(c,1)}_{\mathrm{eh}}$ and $\gamma^{(c,2)}_{\mathrm{eh}}$ and rewrite
$\Gamma_\mathrm{coh}(V)=K_\mathrm{coh}\Gamma_\mathrm{iet}(\Omega_{\mathrm{c},1}+\Omega_{\mathrm{c},2},V)$, where
\begin{multline}\label{eq:G_coh_2}
\Gamma_\mathrm{iet}(\Omega_{\mathrm{c},1}+\Omega_{\mathrm{c},2},V)=\\\frac{\gamma^{(c,1)}_{\mathrm{eh}}+\gamma^{(c,2)}_{\mathrm{eh}}}{\Omega_{\mathrm{c},1}+\Omega_{\mathrm{c},2}}\frac{\Delta_t}{\Delta_{s}}
F(\Omega_{\mathrm{c},1}+\Omega_{\mathrm{c},2},\;eV),
\end{multline}
and $K_\mathrm{coh}=\gamma_\mathrm{coh}\left(\gamma^{(c,1)}_{\mathrm{eh}}+\gamma^{(c,2)}\right)^{-1}$
is the efficiency of the coherent process.

Equation~(\ref{eq:G_coh_2}) shows that the total coherent phonon excitation rate
takes the form of a single-phonon excitation rate $\Gamma_{\mathrm{iet}}(\Omega,V)$ \cite{Tikhodeev2004},
where the single vibrational frequency is replaced by the sum of two vibrational frequencies.

\end{document}